\documentclass[twoside,english]{article}
\usepackage[T1]{fontenc}
\usepackage[latin9]{inputenc}
\usepackage{geometry}
\usepackage{color}
\usepackage{array}
\usepackage{float}
\usepackage{amsmath}
\usepackage{amssymb}
\usepackage{graphicx}
\usepackage[numbers]{natbib}

\makeatletter

\providecommand{\tabularnewline}{\\}

\makeatother

\usepackage{babel}

\begin{document}
\begin{titlepage}
\begin{center}
\medskip
\begin{centering}
\includegraphics[scale=0.28]{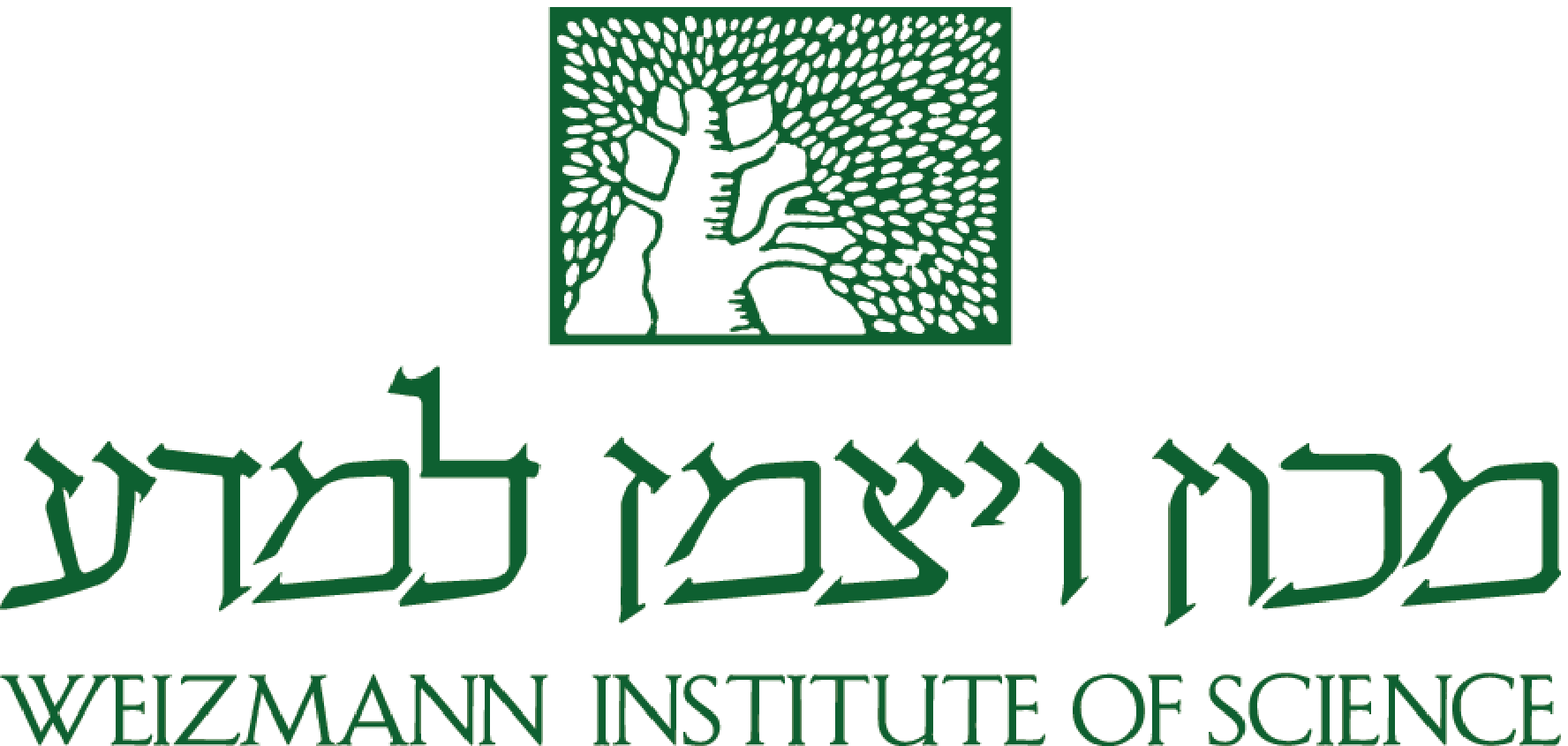}
\par\end{centering}
\bigskip{}
\bigskip{}
\bigskip{}
\bigskip{}
\Large  \textbf{Thesis for the degree Master of Science}\\[0.5cm]
\bigskip{}
\bigskip{}
\bigskip{}
\bigskip{}
\Large \textbf{By Asaf Farhi}\\
\bigskip{}
\bigskip{}
\bigskip{}
\bigskip{}
{ \Large \bfseries Comparing secondary structures of RNA and calculating the free energy of an interior loop using a novel method for calculating free energy differences}\\[0.4cm]
\bigskip{}
\bigskip{}
\bigskip{}
\bigskip{}
\bigskip{}
\bigskip{}
\bigskip{}
\bigskip{}
\Large Advisor: Prof. Eytan Domany\\
\bigskip{}
\bigskip{}
\bigskip{}
\bigskip{}
{\large April 2011}
\end{center}
\end{titlepage}

\section*{Acknowledgements}
\let\thefootnote\relax\footnote{asaf.farhi@gmail.com}
Firstly, I want to thank my supervisor Prof. Eytan Domany. I would
like to thank him for helping with identifying the upcoming obstacles
in the project with his intuition and for introducing me the interesting
and skillful people I had the honor to work with. I appreciate him
for being flexible with the time frames as the project was highly
demanding, and for investing time in reading the thesis and helping
with his exceptional skill for explaining things.

I want to thank Dr. Michael Bon for introducing me the world of RNA
(I used to call him {}``Doctor of RNAs'') and sharing his vast knowledge
in biology, chemistry, physics and computer science. Next, I want
to thank Dr. Guy Hed for the fruitful collaboration and his contribution
in the Monte Carlo simulations and the physics.

I also want to thank Prof. Nestor Caticha for helping with his vast
experience in Monte Carlo simulations.

I want to mention that I really enjoyed working at the Weizmann institute
that provided an excellent environment for scientific research. Specifically
I want to thank to Michal Shoval, Perla Zalcberg, Rachel Goldman and
Yossi Drier that are to be appreciated for their dedication. I want
to thank my group for the social environment and my family for the
support.

There is a provisional patent pending that includes part of the content of this thesis.

\cleardoublepage{}

\section*{Abstract}

The thesis consists of two projects. In the first project, we present
a software that analyses RNA secondary structures and compares them.
The goal of this software is to find the differences between two secondary
structures (experimental or predicted) in order to improve or compare
algorithms for predicting secondary structures. Then, a comparison
between secondary structures predicted by the Vienna package to those
found experimentally is presented and cases in which there exists
a difference between the prediction and the experimental structure
are identified. As the differences originate mainly from faces and
hydrogen bonds that are not allowed by the Vienna package, it is suggested
that prediction may be improved by integrating them into the software.

In the second project we calculate the free energy of an interior
loop using Monte-Carlo simulation. We first present a semi-coarse
grained model for interior loops of RNA, and the energy model for
the different interactions. We then introduce the Monte-Carlo simulations
and the method of Parallel Tempering which enables good sampling of
configuration space by simulating a system simultaneously at several
temperatures. Next we present Thermodynamic Integration, which is
a method for calculating free energy differences. In this method simulations
done at several values of a parameter $\lambda$ are used to yield
the free energy difference in the form of an integral over $\lambda$.
Hence, if simulation at each value of $\lambda$ necessitates parallel
tempering, one has to simulate the system at a set of $\lambda$\textit{
and $T$ }values. We introduce a method that calculates the free energy
significantly faster since we need to use only one parameter, $T$.
To implement this method, we had to reach a regime in which the partition
functions of the two systems are equal, which isn't satisfied for
systems that have different entropies in the high temperature limit,
so a solution to this problem had to be found. Free energy values
calculated for various interior loops are shown, and may, if verified
or with a more realistic modelling, be integrated into the Vienna
package and supply an alternative to the experiments done. More applications
of this method are also suggested.

\cleardoublepage{}

\tableofcontents{}

\cleardoublepage{}

\section{Structural analysis and comparison of RNA secondary structures}

\subsection{Introduction}

In this section, the components of the secondary structures of RNA
will be defined. Then, a software that analyzes these components and
compares secondary structures, will be introduced. The main goal of
the software is to compare two secondary structures and find the differences
between them in order to improve algorithms of prediction or to compare
two algorithms for secondary structure prediction. The software can
be further used to detect elements that are repeated in a certain
group of RNAs and suggest a possible functionality.

\subsection{Secondary structure components}

By convention, the beginning of the RNA molecule is marked by 5' and
the end is marked by 3'. The N nucleotides are referred to as vertices
and the N-1 covalent bonds between consecutive nucleotides are referred
to as exterior edges. Base pairing via Hydrogen bonds is represented
by a line segment called interior edge. The pairings can be between
the following pairs: G-C, A-U and G-U. The entire collection of edges
and vertices is called a graph. An admissible secondary structure
is defined as a graph whose interior edges don't intersect or cross
each other (such crossings are called psuedo-knots).

\begin{figure}[H]
\caption{\label{fig:Illustration-of-the-faces}Illustration of the different
faces}

\medskip{}

~~~~~~~~~~~~~~~~~\includegraphics[scale=1]{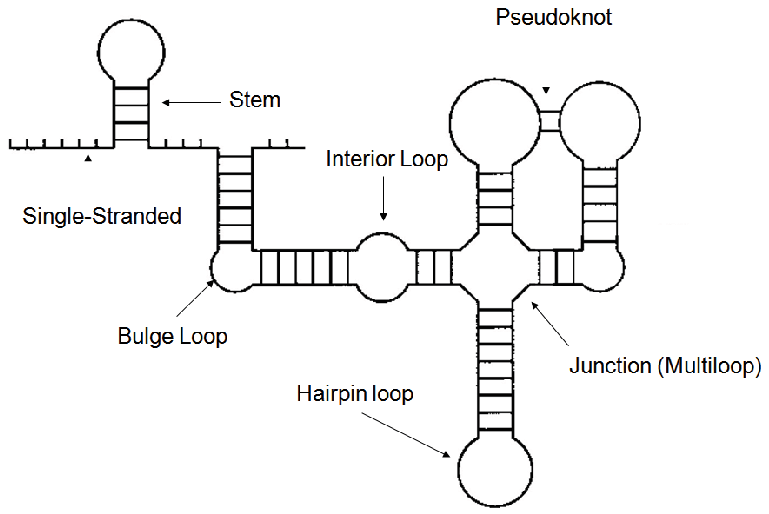}

\end{figure}

A face of a graph is defined to be a planar region, bounded on all
sides by edges. A face with a single interior edge is called hairpin
loop (see example in Fig \ref{fig:Typical-secondary-structure}).
Faces with two interior edges are classified into 3 groups. If the
interior edges are separated by single exterior edges on both sides,
the face is called a stacking region. If they are separated by a single
exterior edge on one side, but by more on the other side, the face
is called a bulge loop. Otherwise the face is referred to as an interior
loop. A face with three or more interior edges is referred to as a
multi-loop or bifurcation loop (see Fig. \ref{fig:Illustration-of-the-faces}).

The word hairpin refers to a structure or substructure, whose faces
are consecutive region of stacking regions, bulges and interior loops,
ending with a hairpin loop \citep{zuker1981optimal}.

\subsection{The software}

The software's main tasks are to re-generate the structure from the
output that represents the hydrogen bond, to analyse it and to compare
it with another structure of similar sequence. The software is suited
for the output of the Vienna package in which a matched pair is represented
by parentheses (example is presented in Figure \ref{fig:Tree-rep}),
and can be easily adopted to suit the output that presents the paired
bases, which is also common. It was programmed in c++ in Unix environment
(Unix compiler), using Eclipse.

\subsection{Generation of the secondary structure}

\begin{figure}[H]
\caption{\label{fig:Tree-rep}An example of a secondary structure $\left(a\right)$
and its parenthesis $\left(b\right)$ and tree representations $\left(c\right)$}

\begin{raggedright}
\medskip{}
$\left(a\right)$~~~~~~~~~~~~~~~~~~~~~~~~~~~~~~~~~~~~~~~~~~~~~~~~~~~~~~
\par\end{raggedright}

\begin{centering}
\includegraphics[scale=1]{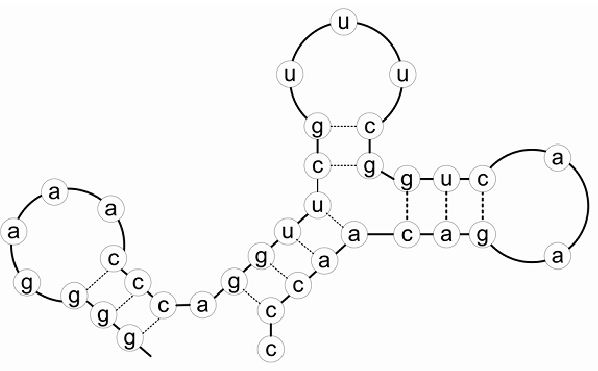}
\par\end{centering}

\begin{centering}
\smallskip{}

\par\end{centering}

\begin{raggedright}
$\left(b\right)$
\par\end{raggedright}

\begin{centering}
(((....))).((((((...))(((..))))))).
\par\end{centering}

\begin{centering}
\smallskip{}

\par\end{centering}

\begin{raggedright}
$\left(c\right)$
\par\end{raggedright}

\begin{centering}
~~~~~~~~~~~~~~~~~~~~~~~~~~structure
\par\end{centering}

\begin{centering}
~~~~~~~~~~~~~~~~~~~~~~~~~~~$\swarrow\searrow$
\par\end{centering}

\begin{centering}
~~~~~~~~~~~~~~~~~structure~~hairpin structure~
\par\end{centering}

\begin{centering}
~~~~~~~~~~~~~~~~~~$\swarrow\,\,\,\,\,\,\,\,\searrow$
\par\end{centering}

\centering{}hairpin structure hairpin structure
\end{figure}

The software uses tree representation (example is presented in Figure
\ref{fig:Tree-rep}) for the structures, which is used for secondary
structures and assumes no psuedo-knots. The software divides the structure
into substructures, which include a sequence of faces with 2 interior
edges and a hairpin loop or a multiloop as their last face. Each substructure
may contain sons, which are substructures, that are connected to it
via a multiloop. In the case of a multiloop, it contains the first
pair of the multiloop, and the sons contain as their first pair the
other pairs of the multiloop.

In the process of identifying the tree representation the faces are
detected, classified and stored in each sub-structure (the pairs are
also stored in a different list). It is here to mention that the software
also stores in each face the unbound nucleotides. A detailed description
of the algorithm is given in Appendix \ref{sub:Description-of-the-algorithm}.
\begin{itemize}
\item The functionality of the software in generating structures was verified
on five RNA outputs.
\end{itemize}

\subsection{Structure comparison}

In order to compare two structures of the same RNA, the faces were
compared. They were divided into three groups: \textit{identical},
meaning that they have the same structure, nucleotide sequences and
\textcolor{black}{nucleotide indexes. }\textit{\textcolor{black}{Similar}}\textcolor{black}{,
meaning that they have the same structure, nucleotide sequence and
not the same nucleotide indexes. And }\textit{\textcolor{black}{different}}\textcolor{black}{,
meaning that it doesn}'t fall into the two categories mentioned (A
detailed description of the algorithm is given in Appendix \ref{sub:Description-of-the-algorithm-comparing}).
This division enables us, in case of different structures, to identify
the list of faces that have different free energy values. Thus, faces
which are suspected of having incorrect energy values may be focused
on. The functionality of this comparison was checked on the structures
of the RNAs: PDB\_00030, PDB\_00213, PDB\_00394 and yielded the expected
results.

\subsection{Applications}

\begin{figure}[H]
\caption{\label{fig:Typical-secondary-structure}Typical secondary structure
of tRNA}

\centering{}\includegraphics[scale=0.75]{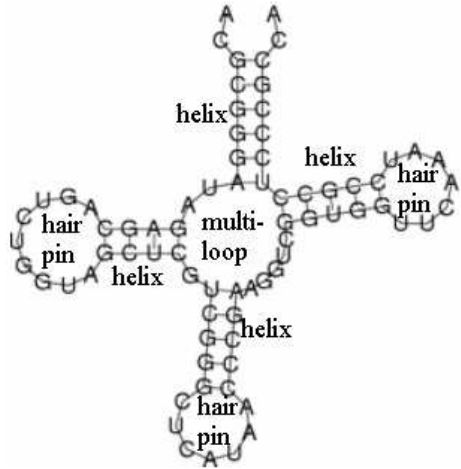}
\end{figure}

The main uses of the structural analysis can be the following:
\begin{itemize}
\item Comparison between structures of minimum free energy, found by different
algorithms, and thus trying to extract in which cases they yield different
results. It can be used to compare the mfe (minimum free energy) structures
predicted by the Vienna package and by other algorithms that have
different free energy parameters. When the structures are different,
the software can output the faces predicted by both and the origin
of the difference can be estimated.
\item Comparison between structures of minimum free energy of a certain
algorithm and the experimental results. One can assess which are the
cases where differences exist. Here again, the origin of difference
can be identified and the parameters of the algorithm can be corrected
accordingly.
\item Identification of structural motives (found by analyzing the structures
predicted by the algorithms) that are common in a group of RNAs, which
may lead to classification and possibl\textcolor{black}{e functionality.
It c}an be also used to associate RNA to a group of RNAs once its
structural motives are known. An example for such structural similarity
is the family of tRNA that includes a helix, which leads to a multiloop
which is followed by three helices that end with an hairpin loop (typical
secondary structure of tRNA is shown in Fig \ref{fig:Typical-secondary-structure}).
A possible application, can be the PASRs (promoter associated small
RNAs) that are transcribed from the promoter's regions. These PASRs
bind at the promoter, and reduce the expression of the associated
gene. The secondary structure of PASRs can be predicted using existing
software. Then, using this software, structural motives that are common
can be detected and a structural possible classification, and even
a functional mechanism can be estimated \citep{fejes2009post}.
\end{itemize}

\begin{itemize}
\item Identification of certain sequences that are repeated in a group of
RNAs, especially in unpaired nucleotides, which may provide clues
about their function\textcolor{black}{{} (can be done using experimental
or predicted secondary structures). An example of repeated sequences
is the selenoprotein mRNA sequences. These mRNAs include the UGA codon
which has a dual function. It signals both the termination of protein
synthesis and incorporation of the amino acid selenocysteine. Decoding
of selenoprotein translation requires the SECIS element, a stem-loop
motif in the 3'-UTR of the mRNA, carrying short or large apical loop.
SECIS elements} contain an Adenosine that precedes the quartet of
non\textendash{}Watson-Crick base pairs, a UGA\_GA motif in the quartet,
and two adenosines in the apical loop or bulge \citep{kryukov2003characterization}\citep{fagegaltier2000structural}.
\end{itemize}

\subsection{Initial results}

The following RNA sequences: PDB 30,32,45,213,394,547,584,750,1236
were compared (experimental structures versus Vienna package results)
to get initial results for the differences between the structures.

It seems that two of the main sources of difference between the experimental
structures and the ones predicted by the algorithm were the following:
\begin{itemize}
\item Bonds that appeared in the experimental structures and weren't allowed
by the algorithms (Vienna package and others), like A-A U-C G-A G-G.
\item Hairpin Loops, containing less than 4 exterior edges (3 in the tested
sequences) were included in the experimental structures and aren't
allowed by the algorithms.
\end{itemize}
It was seen that in some cases (PDB 32,213,394,547,750,1236), the
differences were local, meaning that there were identical faces in
a substructure, then different faces, and then an identical pair which
terminates the local region of difference.

\label{A-A-G-A}Looking in the literature about the pairs which were
detected in the experimental structures checked, and weren't allowed
by the algorithms, it seems that these pairs also show up in other
cases in RNA structures. Such an example is G-A/A-G pairing found
in SECIS element in the 3' untranslated region of Se-protein mRNA
and in the functional si\textcolor{black}{te of the hammerhead ribozyme.
The measured melting temperature values of RNA oligonucleotides having
this pair showed an intermediate value, between that of the Watson-Crick
base pairs and that of non-base pairs \citep{ito2004stability}\citep{ito2001evidence}.
Although rare, A-A and G-G pairs also occur in some cases. It was
seen that a single G-G pair has a stabilizing effect and can stabilize
an internal loop \citep{morse1995purine} and that a group of Adenine
bases can have interactions with each other in a bulge \citep{schneiderwater2}.
Another study shows that the water-mediated UC pair is a structurally
autonomous} building block of the RNA \citep{schneiderwater}. Thus
it seems that the structure of the RNA may be affected also by interactions
that are less dominant.

The tetraloops (Hairpin Loops, containing 3 exterior edges) that were
seen in the experimental data, seem to belong to the three types of
tetraloops that are common in RNA: GNRA,UNCG and CUUG (R=purine, N=every
nucleotide). Meaning that the algorithms exclude these types of faces
although if formed they may lead to a lower total free energy.

Since the total number of different faces may be large, it is possible
to detect regions of difference and output local differences. Thus
differences containing smaller number of faces can be detected, and
tendencies may be pointed out. It can be done by detecting a sequence
of identical faces followed by a sequence of different faces that
ends with an identical pair. Similar faces that belong to that region
can also be detected, and the number of different faces can be further
reduced.

The local differences that were found are the following: (it should
be noted here that the solid and dashed lines represent covalent and
hydrogen bonds respectively and that the arrows point to the experimental
data)

\begin{figure}[H]
\caption{local differences for PDB 32}

\bigskip{}

Experimental structure~~~Vienna package prediction

\medskip{}

(a)\medskip{}

\includegraphics[scale=0.4]{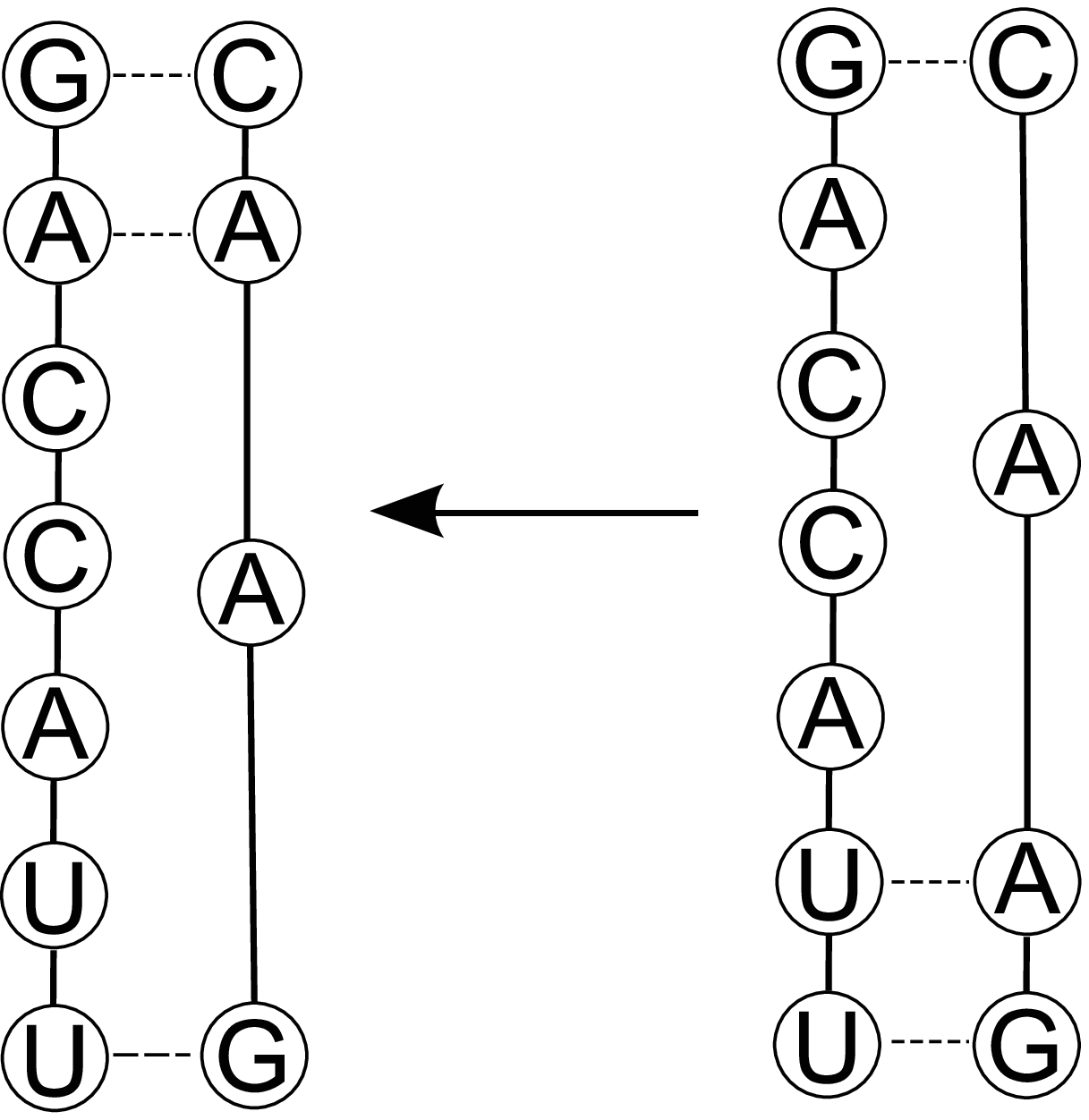}

\medskip{}

(b)

\medskip{}

\includegraphics[scale=0.4]{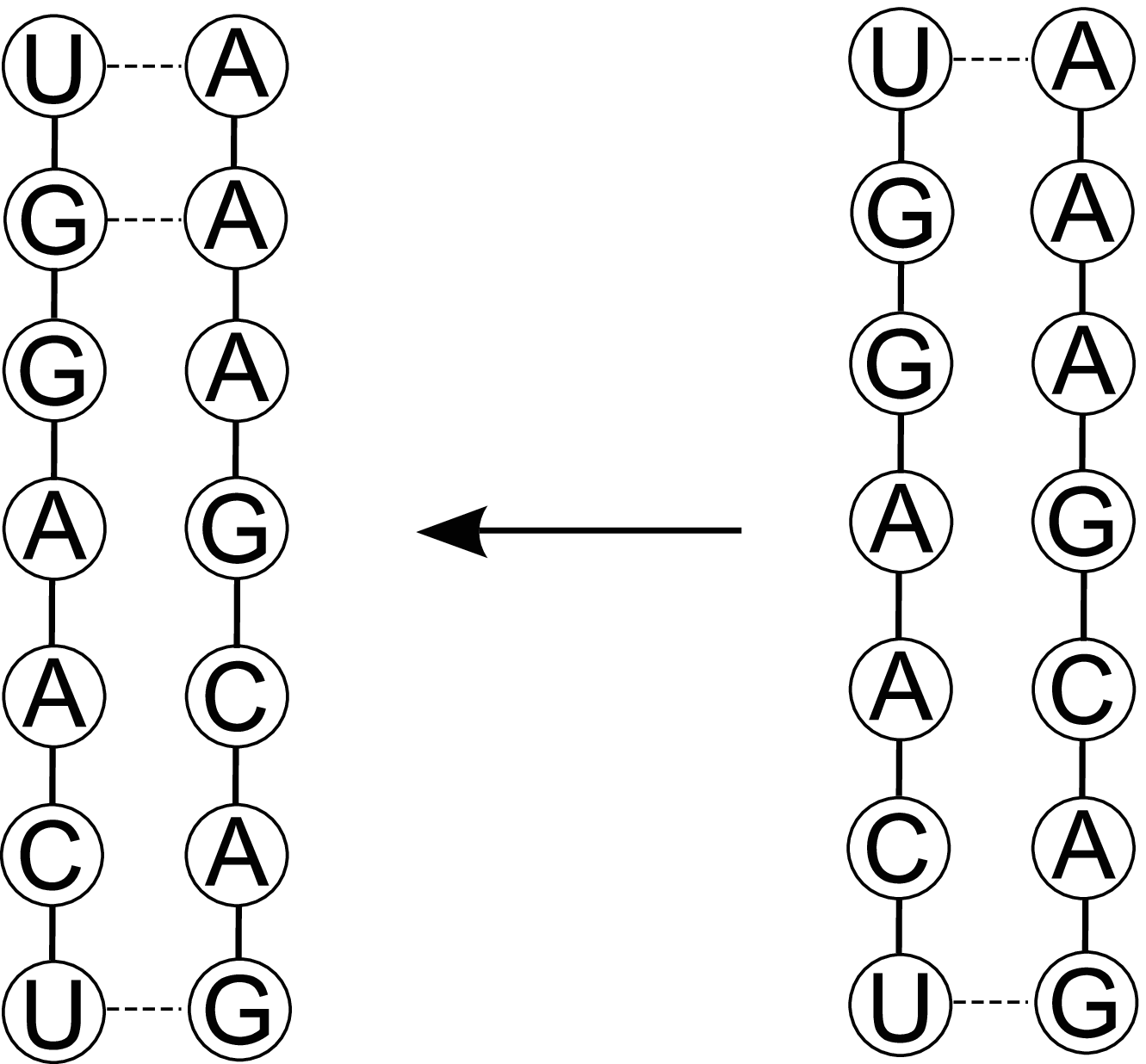}

\medskip{}

(c) Hairpin loop

\medskip{}

\includegraphics[scale=0.4]{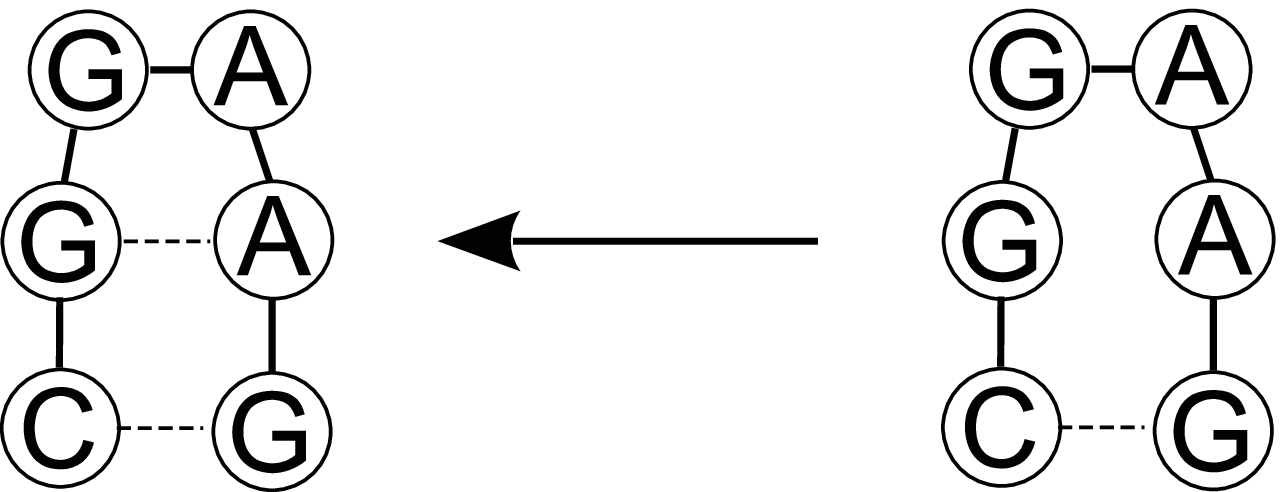}
\end{figure}

\begin{figure}[H]
\caption{local differences for PDB 213}

\bigskip{}

Experimental structure~~~Vienna package prediction

\medskip{}

(a)\medskip{}

\includegraphics[scale=0.4]{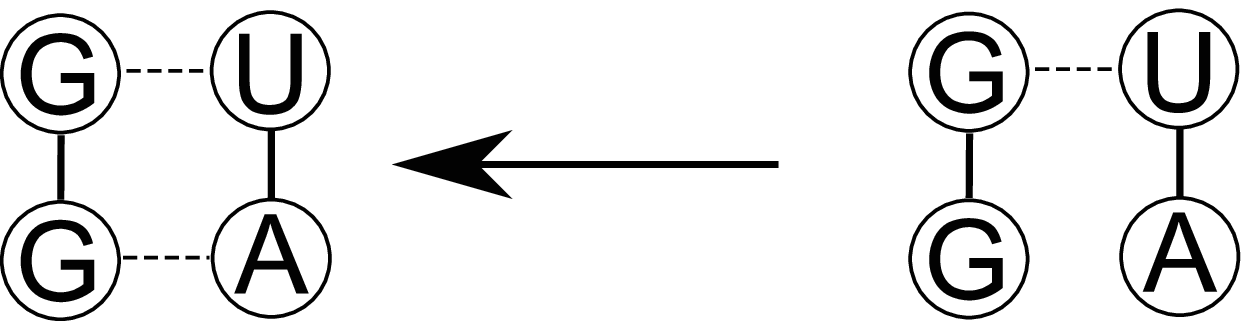}\medskip{}

(b)\medskip{}

\includegraphics[scale=0.4]{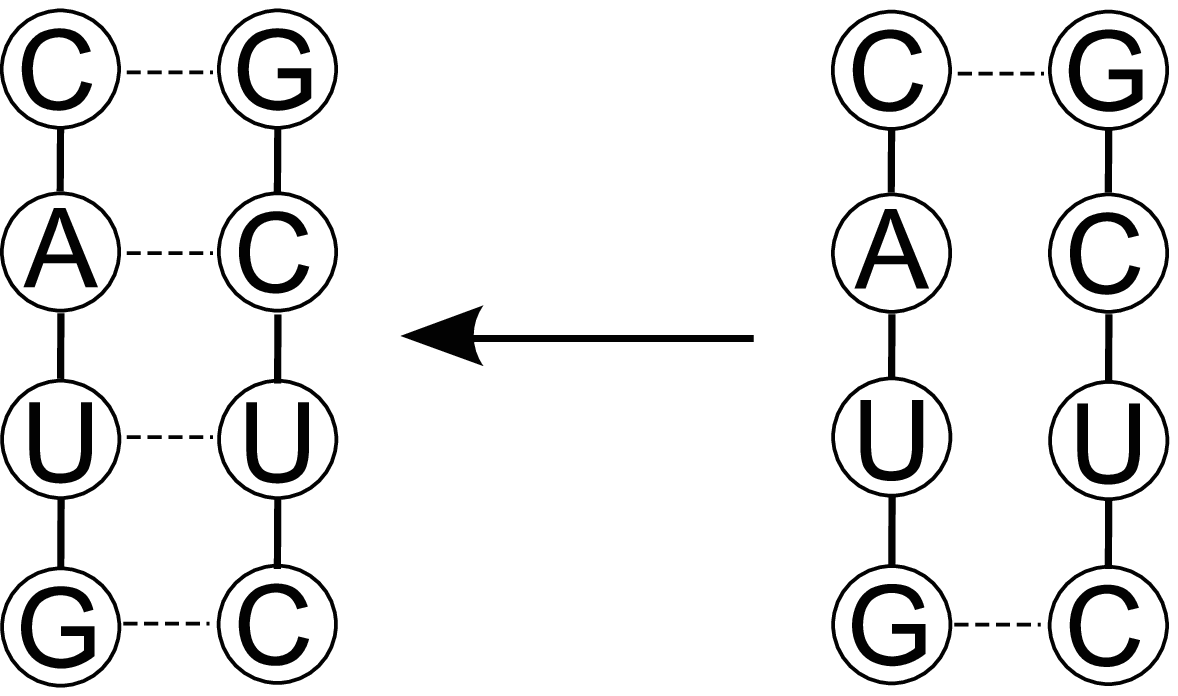}\medskip{}

(c) Hairpin loop\medskip{}

\includegraphics[scale=0.4]{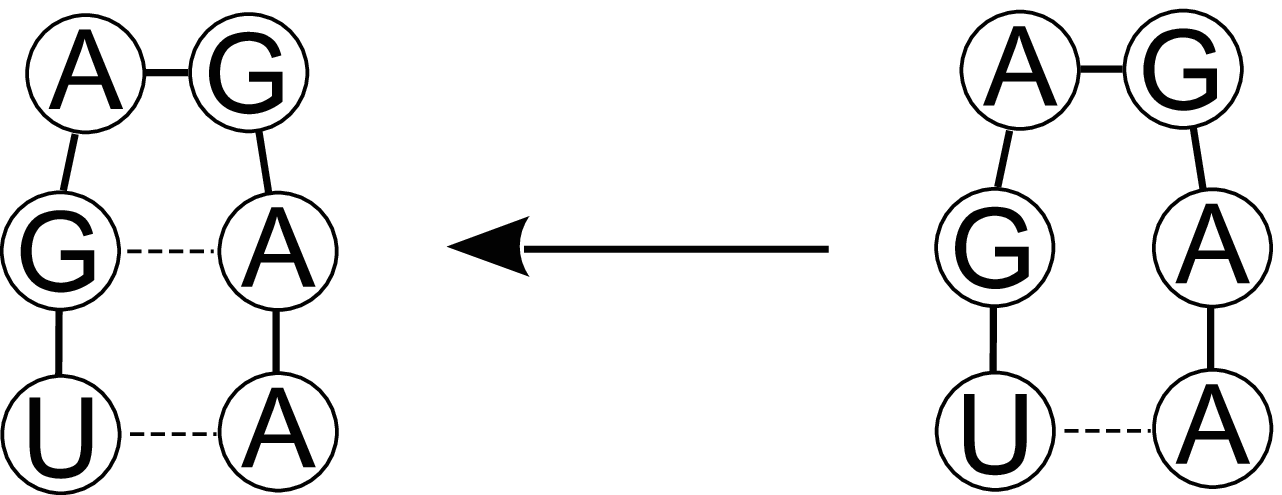}\medskip{}

(d) Hairpin loop\medskip{}

\includegraphics[scale=0.4]{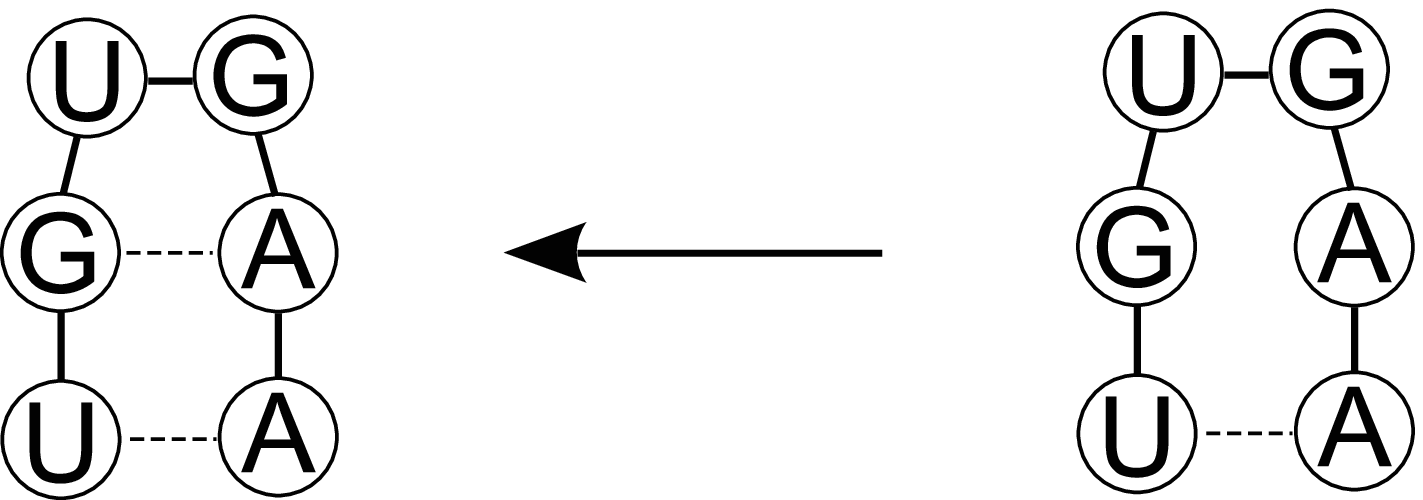}\medskip{}

(e)\medskip{}

\includegraphics[scale=0.4]{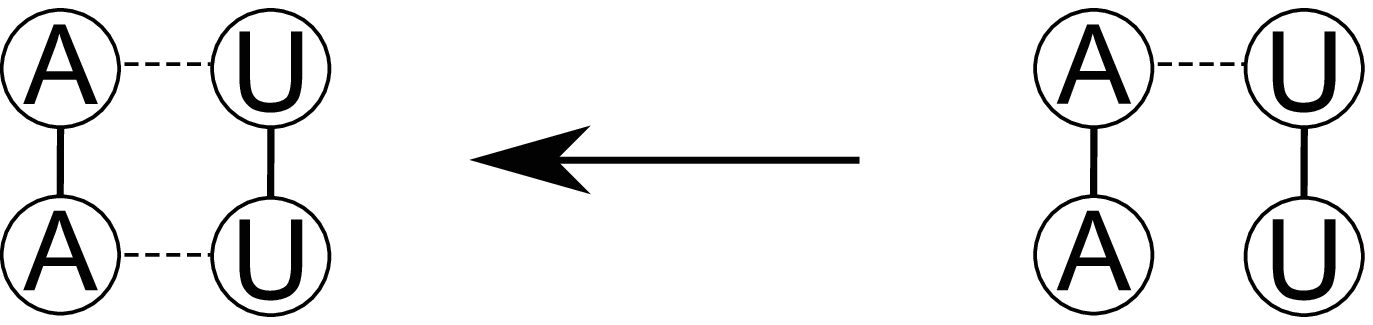}
\end{figure}

~~~~~~

\begin{figure}[H]
\caption{local differences for PDB 394}

\bigskip{}

Experimental structure~~~Vienna package prediction

\medskip{}

(a) Hairpin loop\medskip{}

\includegraphics[scale=0.4]{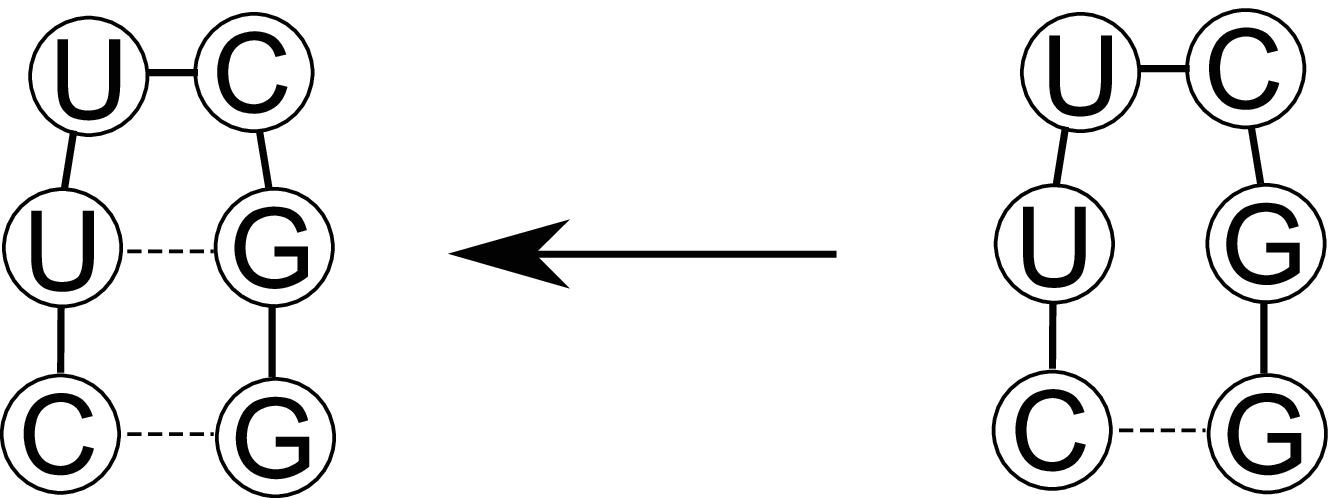}

\medskip{}

(b) Multiloop

\includegraphics[scale=0.4]{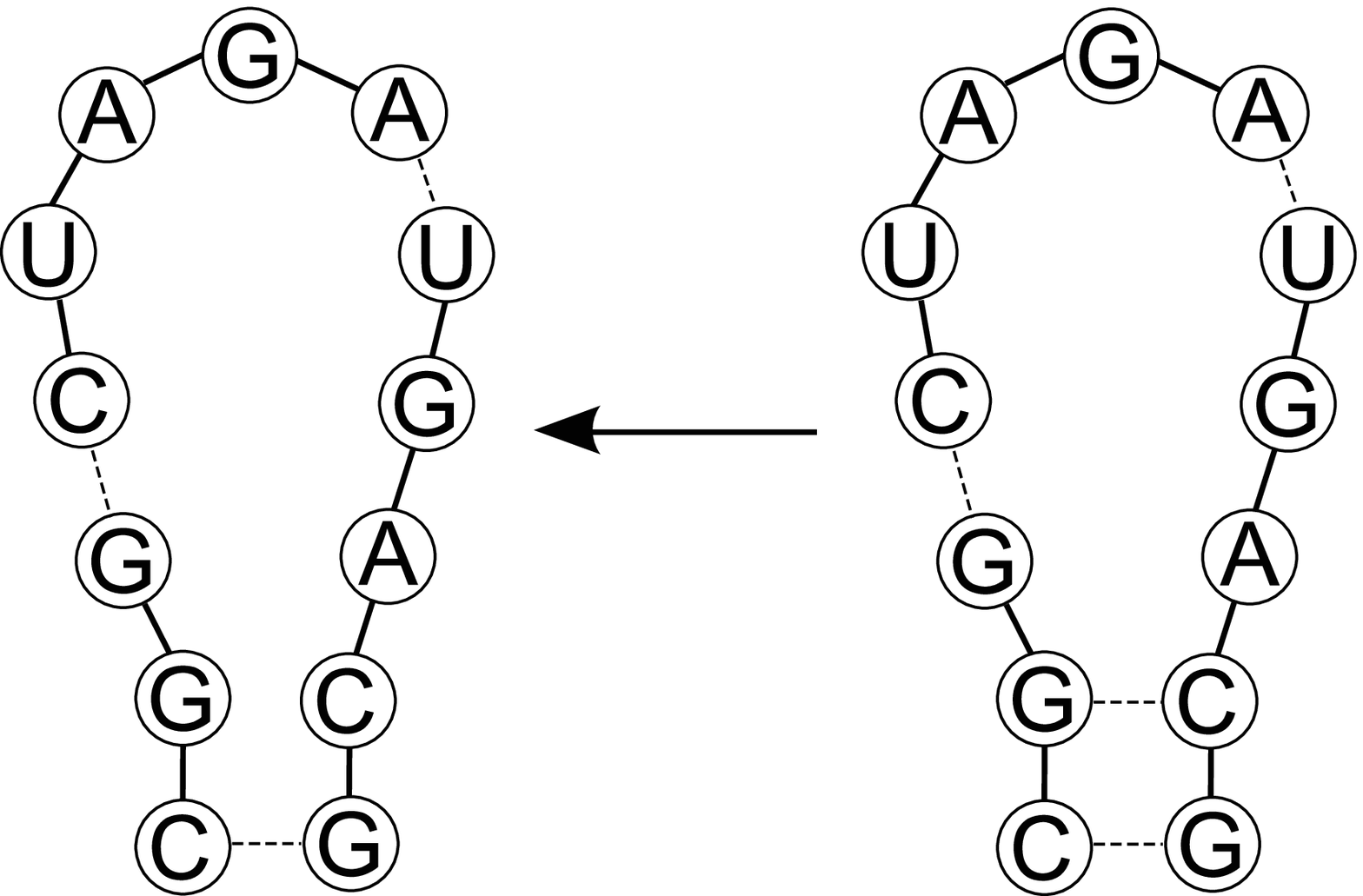}
\end{figure}

\subsection{Discussion}

It can be concluded that the energy parameters of the algorithms give
good predictions for the structures, and in most cases the general
structure is well predicted. However, if we assume that the total
free energy of a structure is the sum of the free energies of the
faces, there are still faces whose free energy parameters aren't estimated
well enough. Local differences may lead to more specific cases of
mismatch, and thus help us focus on the faces that still have to be
investigated.

Certain differences between the structure of PDB 394 found experimentally
and the one predicted by the Vienna package may suggest that the free
energy of a multiloop may vary significantly by its extension by 2
nucleotides. (the multiloop was extended rather than form a G-C hydrogen
bond as suggested by the Vienna package). This may be due to the entropic
loss caused by its reduction by two nucleotides.

Since the differences mainly originate from faces and hydrogen bonds
that are not allowed by the Vienna package it is suggested that prediction
may be improved by integrating parameters of these faces into the
software and predicting unallowed hydrogen bonds according to the
relevant face respectively. This may lead to a more accurate and detailed
description of the pairs in the secondary structure prediction.

\cleardoublepage{}

\section{Calculation of free energy of an interior loop}

\subsection{Introduction}

In the last decade, RNA has gained much attention, due the discovery
that it fulfills a considerable diversity of biological tasks, including
enzymatic and regulatory functions. Since RNA structure is highly
related to its functionality, its of a top priority to predict and
analyse secondary and tertiary RNA structures.

The most common software that predict secondary structure are the
Vienna package/Mfold . These software assume that the total free energy
of a secondary structure is the sum of the free energies of its faces
(planar regions, bounded on all sides by covalent or hydrogen bonds
between nucleotides) \citep{zuker1981optimal}. These free energy
parameters are extracted from experiments in which initial concentrations
of reactors that are RNA segments are changed, and the melting temperature
for the reaction is measured (read Appendix \ref{sub:General-method}
for more details). However, in most of these experiments the heat
capacity was assumed not to depend on the temperature, so they are
to be redone (read Appendix \ref{sub:Finding-enthalpy-and-entropy-values-in-body-temperature}
for more details). Here, we try to check the feasibility of a numerical
calculation of the free energy values of faces, using a Monte-Carlo
simulation. In the case that such a calculation will be feasible,
it might provide an alternative to the experiments that are done,
and supply more specific information, since in the experiments the
faces' free energy values are derived by fitting measured values of
the larger secondary structures in which they are included.

This work is also relevant for calculation of thermodynamic properties
of a complete RNA sequence and its tertiary free energy landscape,
as it can be easily adopted to do so (assuming no pseudoknots). This
is of importance due to the relatively small number of theoretical
works on RNAs as compared to the vast number of studies on proteins.
These theoretical works on RNAs are composed of mostly MD simulations
in which a definition of the coordinate of interest is introduced
in advance, in order to investigate the free energy landscape. There
are relatively fewer Monte-Carlo simulations that involve the calculation
of thermodynamic properties.

In this work we present how the information gathered in a parallel
tempering procedure (details are in \ref{parallel tempering}) can
be used to calculate the free energy differences. In this calculation
we had to reach equality of partition functions and since we didn't
want to be restricted to systems that have the same high temperature
description, we had to go into imaginary regimes in which the steric
constraints were relaxed. This regime was reached due to the use of
cutoff energy which is also relevant to sampling problems that appear
in the Thermodynamic Integration method \citep{kirkwood1935statistical}\citep{straatsma1991multiconfiguration}\citep{frenkel1996understanding}.

\subsection{Goals}

The goals of this project are the following:
\begin{itemize}
\item Model the RNA and the different interactions between the sites.
\item Perform Monte-Carlo simulation in which the space of states will be
sampled correctly, and that will converge and be consistent.
\item Writing a software that will run in reasonable time frames.
\item Calculation of free energy of internal loops that can, in a more elaborate
model, generate the free energy values used in the Vienna package.
\end{itemize}

\subsection{The Model }

In this project, we modeled and simulated an interior loop composed
of 6 nucleotides - 3 in each strand. In this type of interior loop
the first and last pair are bound in canonical or wobble base pairing,
and the pair in the middle can have in the general case all combinations
of bases, not belonging to the canonical or wobble base pairings.
Specifically in this project we chose to simulate the following interior
loop:

\begin{figure}[H]

\caption{Simulated Interior loop}

\medskip{}

\begin{centering}
\includegraphics{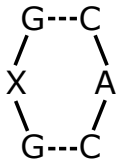}
\par\end{centering}

\end{figure}

where X stands for all types of nucleotides (A,C,G,U).

\subsubsection{Semi-coarse grained model of an Interior loop of RNA}

The model is partly coarse grained and partly in full atomistic detail.
It is coarse-grained in its backbone description and in full atomistic
detail in the description of the nucleobases. This choice of modelling
of the structure is due to the important role the bases play in the
interactions, as the base pairing and the stacking energies are the
dominant ones.

\begin{figure}[H]

\caption{\label{fig:Atomic-numbering-scheme}Atomic numbering scheme and definition
of torsion angles}

\begin{centering}
~~~\includegraphics[scale=1]{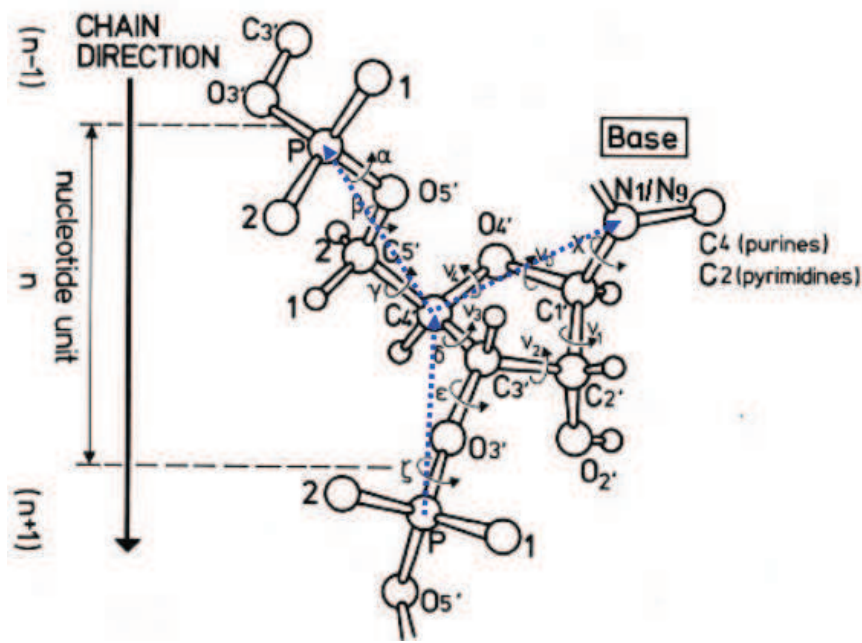}
\par\end{centering}

\end{figure}

\begin{figure}[H]

\caption{\label{fig:The-four-bases}The four bases and their atoms}

\medskip{}

\begin{centering}
~~~~~~~~A ~~~~~~~~~~~~~~~~~~~~~~~~~~~~~~~U
\par\end{centering}

\begin{centering}
\includegraphics[scale=0.11]{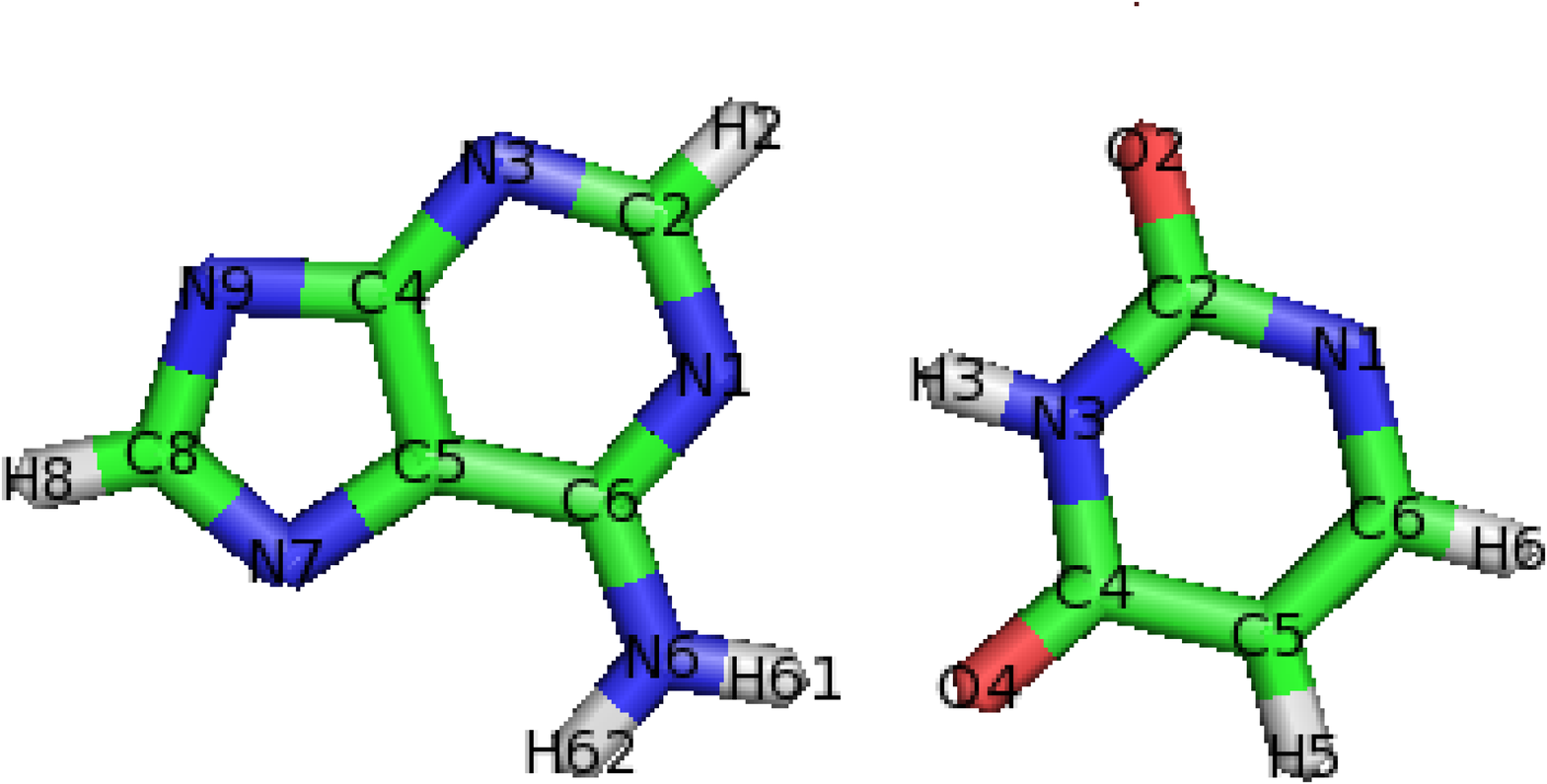}
\par\end{centering}

\begin{centering}
~~~~~~~~~ G~~~~~~~~~~~~~~~~~~~~~~~~~~~~~~~~C
\par\end{centering}

\centering{}\includegraphics[scale=0.11]{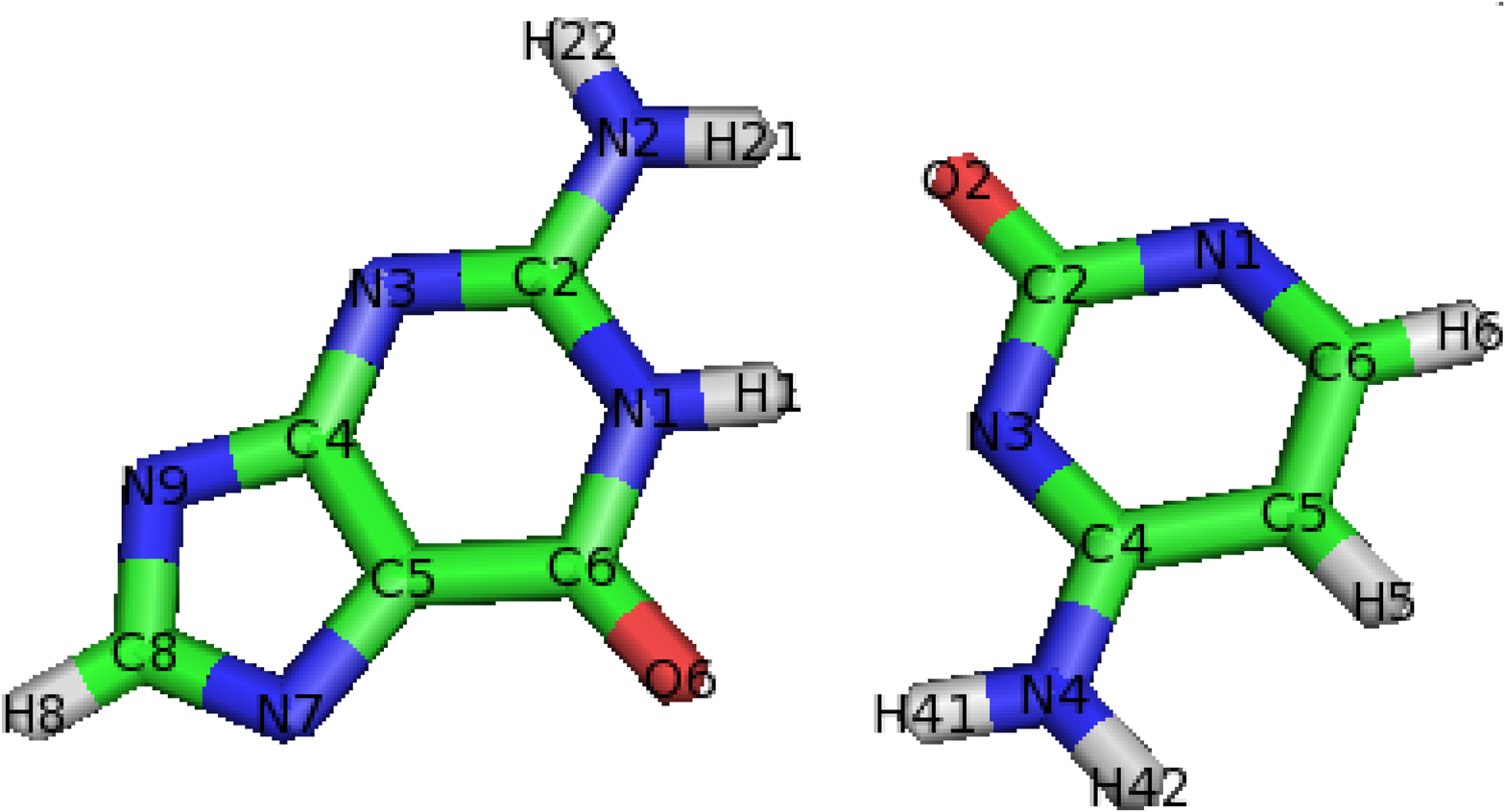}
\end{figure}

The model reduces the complexity of the backbone to two point interaction
sites, one for the phosphate and one for the starting point of the
ribose. The phosphate interaction site was chosen to reside where
the P atom is located (all relevant atoms are presented in fig \ref{fig:Atomic-numbering-scheme}),
and the (beginning of the) ribose interaction site was chosen to reside
where the C4' atom is located. Reasoning for this choice is the fact
that the torsion angles about the two C-O bonds, are preferably in
the trans rotational isomeric state \citep{olson1975configurational},
which means that the P-C4' and C4'-P bond lengths don't vary much.
It should be noted that according to the location of the C4' and P
atoms, the building of the RNA backbone in atomic detail \citep{keating2010semiautomated}
and as a result the ribose can be achieved (The torsion angle $\delta$
determines the configuration of the ribose \citep{murray2003rna}).The
bases were assumed to be rigid planar objects as implied by \citep{saengerp125}.
The modelling of the bases included all the atoms (see Fig. \ref{fig:The-four-bases}
for details), what enables them to form Watson-Crick, and Hoogsteen
base pairing \citep{leontis2001geometric}. The atoms' coordinates
were calculated in the system of coordinates whose two base vectors
reside in the bases' plane (based on the data form \citep{Handbookbiochemistry}).
The bases were assumed to be connected to the C4' atom from the N1
atom in purines and the N9 in pyrimidines. The distances between the
two backbone sites, as well as the distance between the C4' atom and
the N1/N9 atom were chosen to be fixed. This was supported by an analysis
of PDB files performed in which they showed small standard deviations
(it was also reported in \citep{rich1961molecular} that the backbone
bonds defined have lengths of $\sim3.9\overset{\circ}{A}$). The distances
between the interaction sites were chosen to be the ones found in
a specific pdb file (1AFX.pdb), for simplicity. It has to be mentioned
here, that this modelling of the interior loop doesn't necessarily
have less degrees of freedom than a full atomistic model. However,
the treatment in terms of programming time is shorter.

For an interior loop, there are two pairs of canonical/wobble hydrogen
bonds. The pairing is between the first base in one strand and the
last base in the second strand, and between the last base in one strand
and the first base in the second strand. In the model, it was assumed
that the orientation of the paired bases is constant with respect
to each other. Regarding the other bases, they were assumed to have
freedom of movement - the C4'-N bond was free to rotate and the base
was free to rotate with respect to this bond. All the interaction
sites in the backbone were also free to move with the restrictions
that the first pair of bases is constant in space (due to rotation
and translation symmetry), and the last pair of bases is treated like
a rigid object (assumed to have fixed pairing geometry) and free to
move.

\subsubsection{The energy model}

The following energy terms were considered in the model:

\paragraph{Stacking energy}

One of the dominant terms is the stacking energy between the bases,
which originates from dispersive and repulsive interactions. The stacking
energy is calculated between all the atoms in the bases and is represented
by the 6-12 Lennard Jones potential:

\begin{equation}
\mathcal{H}_{\mathrm{stacking}}=\underset{\underset{\alpha\neq\beta}{\alpha,\beta}}{\sum}\underset{i,j}{\sum}\epsilon_{ij}\left[\left(\frac{r_{m_{r_{i\left(\alpha\right)j\left(\beta\right)}}}}{r_{i\left(\alpha\right)j\left(\beta\right)}}\right)^{12}-2\left(\frac{r_{m_{r_{i\left(\alpha\right)j\left(\beta\right)}}}}{r_{i\left(\alpha\right)j\left(\beta\right)}}\right)^{6}\right]
\end{equation}

where $r_{i\left(\alpha\right)j\left(\beta\right)}$ and $r_{m_{r_{i\left(\alpha\right)j\left(\beta\right)}}}$
are the distance between atoms and the distance at which the energy
is minimum respectively ($i\left(\alpha\right)$ is atom $i$ of base
$\alpha$).

It can be easily seen that the minimum energy is $\epsilon$ and that
we get infinite energy at $r\rightarrow0$ and zero energy at $r\rightarrow\infty$.

The distances of minimum energy were chosen from \citep{saengerp40}
and the energy values were chosen from \citep{Van} in order to be
consistent with behavior of bases as reported in \citep{saengerp139}.

The dominant energy terms have distance of minimum energy in the range
$3\overset{\circ}{A}<r_{m_{ij}}<3.4\overset{\circ}{A}$ and minimum
energies in the range $0.15\frac{KCal}{Mole}<\epsilon_{ij}<0.2\frac{KCal}{Mole}$

\paragraph{Base-pairing energy}

Another dominating term is the base-pairing energy that originate
mainly from the hydrogen bonds.

The potential is much more localized and was chosen to be:

\begin{equation}
\mathcal{H}_{\mathrm{bp}}=\underset{i,j}{\sum}h_{bp_{ij}}T_{ij}\,\,\,\,\,\,\,
\end{equation}

where
\begin{equation}
h_{\mathrm{bp_{ij}}}=\varepsilon_{ij}\left[\left(\frac{r_{m_{ij}}}{r_{ij}}\right)^{12}+tgh\left(31\left(\frac{r_{ij}-2}{r_{m_{ij}}}\right)\right)-1-0.8\left(\frac{r_{m_{ij}}}{r_{ij}}\right)^{6}\right]
\end{equation}

and

\begin{equation}
T_{ij=}\left\{ \begin{array}{cc}
cos^{2}\theta_{ij} & h_{\mathrm{bp}}<0\\
1 & h_{\mathrm{bp}}\geq0
\end{array}\right.
\end{equation}

\begin{figure}[H]

\caption{\label{fig:theta} The angle between the acceptor atom, the polar
hydrogen and the donor atom}

\centering{}\includegraphics[scale=1]{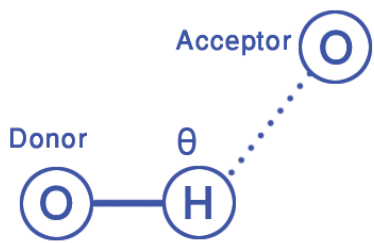}
\end{figure}

$\theta_{ij}$ is the angle between the acceptor atom, the polar hydrogen
and the donor atom (see Fig \ref{fig:theta}), and

$i,j$ are the indexes of the hydrogen and the acceptor respectively.

The term $tgh\left(31\left(\frac{r_{ij}-2}{r_{m_{ij}}}\right)\right)-1$
was introduced to achieve locality in the potential (keeping the energy
zero at infinity), and the term $cos^{2}\theta$ was introduced to
achieve directionality.

The distance of min. energy is $r_{m_{ij}}=1.9\overset{\circ}{A}$
and the min. energies values $\varepsilon_{ij}$ are chosen according
to the acceptor and the electronegativity of the donor. Explicitly,
$\varepsilon_{ij}$ is the product of the difference in electronegativity
between the hydrogen atom and the donor, and a factor that accounts
for the effect of the number of the lone pairs of the acceptor - 1
for N and 2 for O (electrons in the outer shell can participate in
the interaction).

Another proposed potential is with:

\begin{equation}
T_{ij=}\left\{ \begin{array}{cc}
cos^{4}\theta_{ij} & h_{\mathrm{bp}}<0\\
1 & h_{\mathrm{bp}}\geq0
\end{array}\right.
\end{equation}

and with the factor that accounts for the effect of the number of
the lone pairs of the acceptor - 1 for N and 1.15 for O. This choice
is pretty much in accordance with the energy parameters in \citep{vedani1988yeti}.

\paragraph{Electrical energy}

Coulombic interactions are taken into account using the Debye-Huckel
approximation, which is valid for the low-salt physiological conditions.

The Coulombic interactions in the model are between the oxygens next
to the phosphates, and are calculated between phosphates of consecutive
nucleotides (as in \citep{knotts2007coarse}).

\begin{equation}
V_{qq}=\underset{i<j}{\overset{N}{\sum}}\frac{q_{i}q_{j}}{4\pi\varepsilon_{o}\varepsilon_{k}r_{ij}}e^{-r_{ij}/\kappa_{D}}
\end{equation}
~ where
\begin{equation}
\kappa_{D}=\sqrt{\frac{\varepsilon_{0}\varepsilon_{k}RT}{2N_{A}^{2}e_{q}^{2}I}}
\end{equation}
 and $I$ is the ionic strength. We used {[}Na+{]}=50mM ($\kappa_{D}=13.6\overset{\circ}{A}$).

(according to the physiological conditions {[}Na+{]}=150mM {[}Mg++{]}=12mM
($\kappa_{D}=6.835\overset{\circ}{A}$) )

\paragraph{Excluded volume interaction}

The excluded volume terms were introduced in order to obey steric
constraints. They are calculated between all atoms and sites, taking
into account the combinations which aren't included in the stacking
energy and in the base-pairing calculations.

The interaction used is identical to the stacking, shifted up, and
with a cutoff radius above which on this energy is zero.

The term is the following:

\begin{equation}
H_{\mathrm{excluded\, volume_{ij}}}=\left\{ \begin{array}{cc}
\epsilon_{ij}\left[\left(\frac{r_{cutoff}}{r_{ij}}\right)^{12}-2\left(\frac{r_{cutoff}}{r_{ij}}\right)^{6}+1\right] & r<r_{cutoff}\\
0 & r\geq r_{cutoff}
\end{array}\right.
\end{equation}

where we chose $r_{cutoff}=r_{m_{ij}}$.

\newpage{}

In the following figure the typical energies as a function of distance
are presented:

\begin{figure}[H]
\caption{Typical energies as a function of distance}

\begin{centering}
\includegraphics[scale=1]{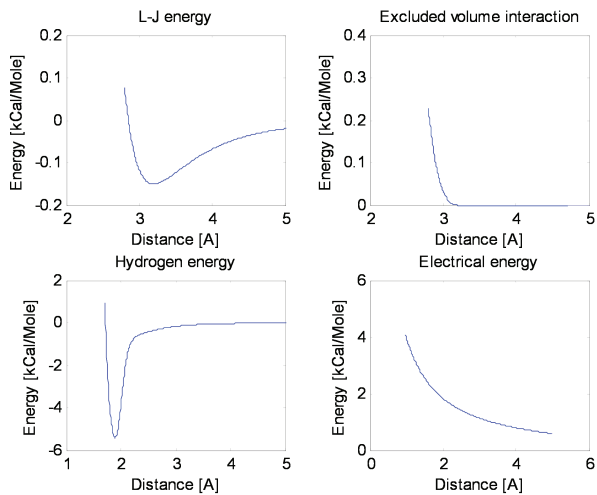}
\par\end{centering}

\end{figure}

\subsection{The Monte-Carlo Simulation and parallel tempering}

In order to to simulate the behavior of the system, the Monte-Carlo
method was used.

It was used with the Metropolis Criterion, according to which a new
configuration is accepted with probability of
\begin{equation}
P_{i\rightarrow j}=min\left\{ 1,e^{-\beta\left(H_{i}-H_{j}\right)}\right\}
\end{equation}

Since the Metropolis Criterion satisfies the detailed balance condition
~(see Appendix 3.4 for details)
\begin{equation}
P_{i}P_{i\rightarrow j}=P_{j}P_{j\rightarrow i},
\end{equation}

all suggested moves were chosen such that a move and its reverse have
the same a-priori probability.

In order to sample all possible states with the model's restrictions,
the following MC moves were used:
\begin{enumerate}
\item External move - a crankshaft move of the backbone. In this move two
sites of the backbone are randomly chosen, and all the sites between
them are rotated by a random angle with respect to the axis that connects
the two chosen sites. The rotation angle was generated according to
a normal distribution with a standard deviation of 7$^{\circ}$.
\item Internal move - random selection of an unpaired nucleotide. Then,
a random selection between the following moves:\end{enumerate}
\begin{itemize}
\item Random rotation of the base with respect to a random vector with N1/N9
held fixed.
\item Random rotation of the C4'-N1/9 vector (bond) and the base, with respect
to a random vector, with C4' held fixed.
\end{itemize}
The rotation angles for both internal moves were generated using normal
distribution with a standard deviation of 5$^{\circ}$.

In order to generate random vectors, $\theta$ was generated with
the distribution of
\begin{equation}
f\left(\theta\right)=\frac{sin\theta}{2}\,,\theta=\left[0,\pi\right]
\end{equation}
 and $\phi$ was generated with a uniform distribution
\begin{equation}
f\left(\phi\right)=\frac{1}{2\pi},\,\phi=\left[0,2\pi\right]
\end{equation}
 Thus, the locations on the unit sphere/the solid angles were generated
with a uniform distribution~
\begin{equation}
f\left(\Omega\right)=\frac{1}{4\pi},\,\Omega=\left[0,4\pi\right]
\end{equation}

In each move, after the locations of the sites are updated, the bond
and torsion angles are calculated (according to \citep{parsons2005practical}).

In order to sample the states of configurations correctly during the
simulation without being trapped in local minima, parallel tempering
has been used \citep{hansmann1997parallel}.

\label{parallel tempering}

The general idea of parallel tempering is to simulate M replicas of
the original system of interest, each replica at a different temperature.
The high temperature systems are generally able to sample large volumes
of phase space, whereas low temperatures systems, whilst having precise
sampling in a local region of phase space, may become trapped in local
energy minima during the timescale of a typical simulation. Parallel
tempering achieves good sampling by allowing the systems at different
(usually adjacent) temperatures to exchange complete configurations.
Thus, the inclusion of higher temperatures ensures that the lower
temperature systems can access a representative set of low temperature
regions of phase space. The M systems can be considered as one artificial
system that contains M non-interacting replicas. We can define the
state of the artificial system as $\mathcal{C}=\left\{ C_{1},C_{2},...,C_{M}\right\} $,
where $C_{i}$ is the state of replica $i$ with energy $H_{i}=H\left(C_{i}\right)$
. It can then be seen, that the probability of switching adjacent
configurations according to the Metropolis criterion is
\begin{equation}
P\left(\mathcal{C}\rightarrow\mathcal{C}'\right)=min\left\{ 1,\frac{e^{-\beta_{i}H_{i+1}-\beta_{i+1}H_{i}}}{e^{-\beta_{i}H_{i}-\beta_{i+1}H_{i+1}}}\right\} =min\left\{ 1,e^{+\left(\beta_{i+1}-\beta_{i}\right)\left(H_{i+1}-H_{i}\right)}\right\} =min\left\{ 1,e^{\Delta}\right\}
\end{equation}

To ensure that an exchange of the conformation between two copies
will happen with sufficient probability, $\Delta$ has to be on the
order of one. If we approximate $H_{i}$ as the average energy of
system $i$, we can write:
\begin{equation}
\Delta\approx\left(\triangle\beta\right)^{2}\frac{\partial U}{\partial\beta}
\end{equation}

where $\Delta\beta=\beta_{i+1}-\beta_{i}$

Since the average energy grows roughly linearly with the number of
residues $N$, $\Delta\beta$ should be of the order of $\frac{1}{\sqrt{N}}$
\citep{earl2005parallel}.

In our case $N\approx70$, so we need approximately 8-9 different
temperatures, that have to be spread up to a temperature at which
the dominating interactions will have a mild effect on the system's
behavior.

\subsubsection{Ratios between the different MC moves}
\begin{itemize}
\item For each external move, 90 internal moves are performed.
\item For each configuration exchange attempt, 10 external moves are performed.
\end{itemize}

\subsection{Parallel tempering and the calculation of free energy differences}

\subsubsection{Introduction}

\begin{figure}[H]
\caption{\label{fig:interior_loops}$\tilde{C}$ and $\tilde{G}$}

\smallskip{}

\centering{}\includegraphics{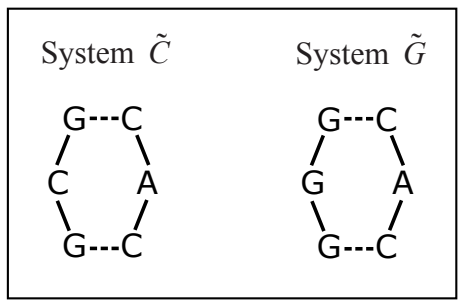}
\end{figure}

Our goal is to find the free energy difference between two systems:
two interior loops, which differ in a base. Specifically we will calculate
the free energy difference between the interior loops $\tilde{G}$,
whose second base is G, and $\tilde{C}$, with the base C, as shown
in Fig \ref{fig:interior_loops}.

We denote this difference by $\triangle F_{\tilde{C}\rightarrow\tilde{G}}\left(\beta_{1}\right)=F_{\tilde{G}}(\beta_{1})-F_{\tilde{C}}(\beta_{1})$.

The state of the bases $G,C$ (the coordinates of all their atoms)
for a given location of the N1/N9 atom of the base (see Fig \ref{fig:The-four-bases}
for details) is determined by three angles, since the bases are assumed
to be rigid planar objects. Therefore the same degrees of freedom
are associated with these different bases, and since the two systems
$\tilde{G},\tilde{C}$ differ just in the bases, the two systems have
the same degrees of freedom.

The difference of free energy between any two systems that have the
same degrees of freedom can, in principle, be calculated by the Thermodynamic
Integration (ThI) method, as described in section \ref{sub:Thermodynamic-integration}.
ThI requires running simulations of different instances (several values
of a parameter $\lambda_{i},i=1,2,...m$) of a hybrid $\tilde{G}-\tilde{C}$
system.

These systems have a rugged energy landscape (with many local minima),
which means that the decorrelation time in a {}``naive'' Monte-Carlo
simulation are exponentially long. One of the widely used methods
to alleviate the problem of equilibrating systems with such landscapes
is parallel tempering (PT), which we describe below in Section \ref{sub:TI-and-parallel-tempering}.
PT involves simulating the system of interest, in parallel, at several
(inverse) temperatures $\beta_{i},i=1,2,...n$.

The natural way to calculate the free energy difference for our two
systems, $\triangle F_{\tilde{C}\rightarrow\tilde{G}}\left(\beta_{1}\right)$
is, therefore, by ThI, equilibrating $m$ systems, one for each value
of $\lambda_{i}$. For each such equilibration PT is used, at $n$
temperatures, hence all together $nm$ systems are simulated. In section
\ref{sub:The-method} we describe a method, Temperature Integration
(TeI), which allows us to calculate the free energy difference much
more efficiently than the use of ThI with parallel tempering.

\subsubsection{\label{sub:Thermodynamic-integration}Thermodynamic Integration}

Thermodynamic Integration (ThI) is one of the most commonly used methods
for calculating the free energy difference between two systems. It
is described in detail in Appendix 3.5. This method is used to calculate
the free energy difference between two systems with the same degrees
of freedom \citep{frenkel1996understanding,kirkwood1935statistical,straatsma1991multiconfiguration},
by varying a parameter $\lambda$ that interpolates between the two
compared systems. We will use the systems $\tilde{C}$ and $\tilde{G}$
described above to demonstrate application of this general method
in a specific context.

Denote the Hamiltonians of the two systems by $H_{\tilde{C}}(\mathbf{\boldsymbol{\theta}})$
and $H_{\tilde{G}}(\mathbf{\boldsymbol{\theta}})$, where $\mathbf{\boldsymbol{\theta}}$
denotes the coordinates of the system. Noting that the two systems
have the same coordinate space, we define a $\lambda$ - weighted
hybrid system, characterized by the Hamiltonian $H(\lambda):$
\begin{equation}
H(\lambda,\mathbf{\boldsymbol{\theta}})=\lambda H_{\tilde{G}}(\mathbf{\boldsymbol{\theta}})+(1-\lambda)H_{\tilde{C}}(\mathbf{\boldsymbol{\theta}})\label{eq:Hamiltonian}
\end{equation}

As shown in Appendix 3.5, the free energy difference is given by
\begin{equation}
\triangle F_{\tilde{C}\rightarrow\tilde{G}}\left(\beta_{1}\right)=\intop_{0}^{1}\left[\left\langle H_{\tilde{G}}(\mathbf{\boldsymbol{\theta}})\right\rangle _{\lambda}-\left\langle H_{\tilde{C}}(\mathbf{\boldsymbol{\theta}})\right\rangle _{\lambda}\right]d\lambda\label{eq:reg}
\end{equation}
 where $\left\langle X\right\rangle _{\lambda}$ denotes the equilibrium
average of $X$ in the ensemble characterized by $H(\lambda)$. The
expression for $\triangle F_{\tilde{C}\rightarrow\tilde{G}}\left(\beta_{1}\right)$,
written explicitly, takes the form:

\begin{equation}
\triangle F_{\tilde{C}\rightarrow\tilde{G}}\left(\beta_{1}\right)=\intop_{0}^{1}\left\{ \int\left[H_{\tilde{G}}(\mathbf{\boldsymbol{\theta}})-H_{\tilde{C}}(\mathbf{\boldsymbol{\theta}})\right]\frac{e^{-\beta_{1}\left[\lambda H_{\tilde{G}}(\mathbf{\boldsymbol{\theta}})+\left(1-\lambda\right)H_{\tilde{C}}(\mathbf{\boldsymbol{\theta}})\right]}d\mathbf{\boldsymbol{\theta}}}{Z(\lambda)}\right\} d\lambda\label{eq:detailed}
\end{equation}
 with
\begin{equation}
Z(\lambda)=\int e^{-\beta_{1}\left[\lambda H_{\tilde{G}}(\mathbf{\boldsymbol{\theta}})+\left(1-\lambda\right)H_{\tilde{C}}(\mathbf{\boldsymbol{\theta}})\right]}d\mathbf{\boldsymbol{\theta}}
\end{equation}

The integration is performed numerically, with the integrand evaluated
at each one of a set of values of $\lambda$ by Monte Carlo simulations.
As implied by \eqref{eq:detailed}, the two systems are required to
have the same degrees of freedom.

\subsubsection{\label{sub:TI-and-parallel-tempering}Thermodynamic integration with
parallel tempering}

The systems $\tilde{C}$ and $\tilde{G}$, and the $\lambda$-weighted
system have\textcolor{blue}{{} }\textcolor{black}{rugged} energy landscapes
with many local minima. As the decorrelation times grow exponentially
with $\triangle E/kT$ , where $\triangle E$ is the energy barrier
between nearby valleys, equilibration times in these systems can be
rather long. In order to overcome this problem, and obtain the thermodynamic
averages $\left\langle H(\theta)\right\rangle _{\lambda}$, one can
use the Parallel Tempering procedure \citep{hansmann1997parallel,earl2005parallel}.
However, implementing both thermodynamic integration and parallel
tempering yields an unnecessary overhead in running time, as we will
demonstrate.

In order to implement thermodynamic integration we first choose a
set of values $\lambda_{i},~i=1,...m$, that will enable us to have
a good sampling of the function $\left\langle H(\mathbf{\boldsymbol{\theta}})\right\rangle _{\lambda}$
for the integration in eq. \eqref{eq:reg}. Thus, $m$ is related
to the desired precision of the integration.

In principle parallel tempering should be performed for each $\lambda_{i}$,
simulating each of the $m$ $\lambda_{i}$ weighted systems over a
set of temperatures, given by $\beta_{j},~j=1,...n$. Finally, using
the calculated values of $\left\langle H(\mathbf{\boldsymbol{\theta}})\right\rangle _{\lambda_{i}}$
at a temperature of interest $\beta_{1}$, we approximate the integral
of eq. \eqref{eq:reg} by a sum over $m$ terms, to get the free energy
difference $\triangle F_{\tilde{C}\rightarrow\tilde{G}}\left(\beta_{1}\right)$.

\begin{figure}[H]
\caption{\label{fig:grid}Illustration of the grid of values over which the
Monte-Carlo simulations are performed. The lowest values of $\beta$
correspond to some very high finite temperature. }

\centering{}\includegraphics[scale=0.45]{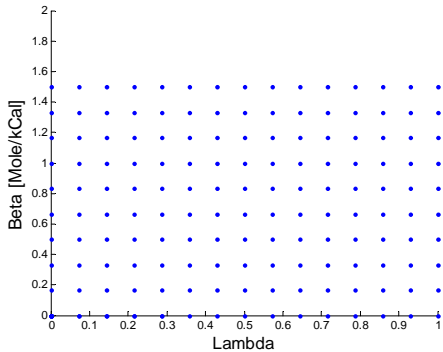}
\end{figure}

The procedure involves running Monte-Carlo simulations over $m$ $\lambda$-values,
and $n$ $\beta$-values, that is, over a grid of $m\times n$ instances
of the\textcolor{black}{{} hybrid} system. An illustration of this grid
is presented in figure \ref{fig:grid}.

\subsubsection{\label{sub:The-method}Calculation of $\Delta F$ by parallel tempering}

We present now a method which uses only simulations obtained in the
process of parallel tempering (skipping the need for simulations at
a set of $\lambda_{i}$ values), performed for the two systems $\tilde{C}$
and $\tilde{G}$, to obtain the free energy difference $\triangle F_{\tilde{C}\rightarrow\tilde{G}}\left(\beta_{1}\right)$.
This method can be applied to any two systems that have the same degrees
of freedom $\boldsymbol{\theta}$.

As $\beta\to0$, the limiting value of the partition function of a
system yields the phase space volume. In particular, if systems $\tilde{C}$
and $\tilde{G}$, which have the same coordinate space $\boldsymbol{\left\{ \theta\right\} }$,
have the same $\beta\to0$ limit, we have
\begin{equation}
Z_{\tilde{G}}\left(\beta\to0\right)=Z_{\tilde{C}}\left(\beta\to0\right)=\int d\boldsymbol{\theta}\,.\label{eq:Z0Z0}
\end{equation}

In this case we can use the following identity to obtain, for a finite
$\beta_{1}$, the difference of the free energies:

\begin{equation}
\ln Z\left(\beta_{1}\right)-\ln Z\left(\beta\to0\right)=\int_{0}^{\beta_{1}}\frac{d\ln Z}{d\beta}d\beta=-\int_{0}^{\beta_{1}}\left\langle H\right\rangle d\beta\,.\label{eq:intH}
\end{equation}

Using equations (\ref{eq:Z0Z0}) and (\ref{eq:intH}) we obtain
\begin{eqnarray}
\triangle F_{\tilde{C}\rightarrow\tilde{G}}\left(\beta_{1}\right) & = & \frac{1}{\beta_{1}}\left[\ln Z_{\tilde{C}}\left(\beta_{1}\right)-\ln Z_{\tilde{G}}\left(\beta_{1}\right)\right]\nonumber \\
 & = & \frac{1}{\beta_{1}}\left[\intop_{0}^{\beta_{1}}\left\langle H_{\tilde{G}}\right\rangle d\beta-\intop_{0}^{\beta_{1}}\left\langle H_{\tilde{C}}\right\rangle d\beta\right]\,.\label{eq:DFCG}
\end{eqnarray}

For each of the systems $\tilde{C}$ and $\tilde{G}$ we estimate
the integrals on the right hand side by parallel tempering, sampling
the system at a series of values $\beta_{1},\ldots,\beta_{n}$. We
choose values such that $\beta_{n}^{-1}$ is much larger than the
internal energy of the system at $\beta_{1}$, so $Z(\beta_{n})\cong Z(\beta\to0)$.

The condition of eq. \eqref{eq:Z0Z0} poses a problem for the particular
systems $\tilde{C}$ and $\tilde{G}$ studied here. The steric excluded
volume interactions are not bound in our model, and hence, the available
phase space volumes for the two systems are not equal at any temperature,
and equation \eqref{eq:Z0Z0} does not hold. In order to satisfy the
condition stated above we had to set a cutoff over the steric interactions,
$E_{\mathrm{cutoff}}$. In the next section we show that our results
do not depend on the choice of $E_{\mathrm{cutoff}}$ and $\beta_{n}$,
as long as $E_{\mathrm{cutoff}}$ is much larger than any typical
interaction energy in the system at $\beta_{1}$, and $\beta_{n}^{-1}\gg E_{\mathrm{cutoff}}$.

\subsubsection{Passing the steric energy barriers by introducing a cutoff energy}

\label{sub:cutoff}

We now tackle the problem of sampling at temperatures which are higher
than all energy terms, by using a cutoff energy in the steric interaction
potential. We denote the cutoff energy and the distance from which
the energy will be set to the cutoff energy by $E_{\mathrm{cutoff}}$
and $r_{\mathrm{cutoff}}$ respectively.

The modified steric potential can be written as follows:

\medskip{}

$\left(\frac{\sigma}{r}\right)^{12}\rightarrow H_{\mathrm{steric}}\left(r\right)=\left\{ \begin{array}{cc}
\left(\frac{\sigma}{r}\right)^{12} & r>r_{\mathrm{cutoff}}\\
E_{\mathrm{cutoff}} & r\leq r_{\mathrm{cutoff}}
\end{array}\right.$

\newpage{}

In the following figure an example of such a steric potential is presented:

\begin{figure}[H]
\caption{Plot of an example of a steric interaction potential that includes
a cutoff energy}

\centering{}\includegraphics[scale=0.37]{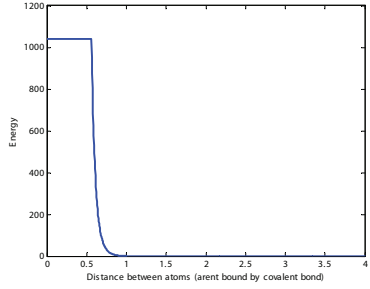}
\end{figure}

The proposed calculation of the free energy difference between the
two systems at the temperature of interest $\beta_{1}$ is legitimate
only if our choice of the cutoff energy has a negligible effect on
the partition function of each of the two systems at $\beta_{1}$.
In addition, the highest temperature used, $\beta_{2}$, must be such
that the equality of the partition functions of the two systems is
satisfied to a good accuracy.

We denote the Hamiltonian with the cutoff energy by $H'$, and write
the requirements stated above explicitly as follows:

\begin{equation}
lnZ_{\tilde{G}}\left(\beta_{1},H\right)\simeq lnZ_{\tilde{G}}\left(\beta_{1},H'\right)
\end{equation}

\begin{equation}
lnZ_{\tilde{C}}\left(\beta_{1},H\right)\simeq lnZ_{\tilde{C}}\left(\beta_{1},H'\right)
\end{equation}

\begin{equation}
lnZ_{\tilde{G}}\left(\beta_{2},H'\right)\simeq lnZ_{\tilde{C}}\left(\beta_{2},H'\right)
\end{equation}

In order for the cutoff to have a negligible effect on the partition
functions at the temperature of interest it has to be set to a value
that satisfies $E_{\mathrm{cut\, off}}\gg kT_{1}$.

As for $\beta_{2}$, if the cutoff energy satisfies $E_{\mathrm{cut\, off}}\ll kT_{2}$,
the systems will have almost equal probability to be in all the regions
of their phase space, including ones in which the molecules entered
the {}``volume'' of one another. Thus, the partition functions of
the two systems will be almost equal.

Hence if these requirements are satisfied one can write:

\begin{equation}
lnZ_{\tilde{C}}\left(\beta_{1},H\right)-lnZ_{\tilde{G}}\left(\beta_{1},H\right)\simeq
\end{equation}

\begin{equation}
lnZ_{\tilde{C}}\left(\beta_{1},H'\right)-lnZ_{\tilde{G}}\left(\beta_{1},H'\right)\simeq
\end{equation}

\begin{equation}
lnZ_{\tilde{G}}\left(\beta_{2},H'\right)-lnZ_{\tilde{G}}\left(\beta_{1},H'\right)-\left[lnZ_{\tilde{C}}\left(\beta_{2},H'\right)-lnZ_{\tilde{C}}\left(\beta_{1},H'\right)\right]
\end{equation}

\newpage{}

Using the identity in eq \eqref{eq:intH}, we can write:

\begin{equation}
\triangle F_{C\rightarrow G}\left(\beta_{1},H\right)=\frac{1}{\beta_{1}}\left[lnZ_{\tilde{C}}\left(\beta_{1},H\right)-lnZ_{\tilde{G}}\left(\beta_{1},H\right)\right]\simeq\frac{1}{\beta_{1}}\left[\intop_{\beta_{2}}^{\beta_{1}}\left\langle H'_{G}\right\rangle d\tilde{\beta}-\intop_{\beta_{2}}^{\beta_{1}}\left\langle H'_{C}\right\rangle d\tilde{\beta}\right]
\end{equation}

So the calculation of free energy will be negligibly affected by the
use of a cutoff energy as long as we fulfill the relevant conditions.

It has to be mentioned that this use of cutoff energy is relevant
also to ThI and similar methods. In these methods if the systems have
different $\beta\to0$ limit, at $\lambda=0,1$ due to the steric
energy terms (represented by $\left(\frac{\sigma}{r}\right)^{12}$)
micro-states may have infinite energy. As a result the internal energy
will be infinite and the sampling of the internal energy will be infeasible.

It can be seen that since the free energy difference between systems
that have the Hamiltonian with the cutoff $H'$ is almost equal to
the one calculated between systems with the Hamiltonian $H$

$\triangle F_{C\rightarrow G}\left(\beta_{1},H\right)\simeq\frac{1}{\beta_{1}}\left[lnZ_{\tilde{C}}\left(\beta_{1},H'\right)-lnZ_{\tilde{G}}\left(\beta_{1},H'\right)\right],$

ThI can be derived for systems with $H'$ and yield almost the same
result.

In conclusion, with the use of cutoff energy in ThI, that enables
us to sample the integrand in all cases, the free energy difference
between any two systems that have the same degrees of freedom can
be calculated.
\begin{itemize}
\item Cutoff energies chosen for the simulation : $E_{cutoff1}=5\left[KCal/Mole\right]$,
$E_{cutoff2}=4.4\left[KCal/Mole\right]$. These cutoff energies satisfy
the conditions mentioned above. Hence, the results of the calculations
of the free energy for the two cutoffs should be similar.
\end{itemize}
\newpage{}

\subsection{The Software and the simulations }

\subsubsection{General description}

The project was written in c/c++ in Unix environment under Eclipse.
It was mostly written in object oriented c++, and at places where
performance was critical it was written in c. The model was divided
into classes that represent its components (Oligonucleotide,Nucleotide,
Base etc.). In cases where the behavior of the class varies according
to its type, derived classes were used (classes that inherit properties
of a base class and implement more properties according to their type).
For example nucleotides were divided into regular nucleotides, first
nucleotide in first strand, last nucleotide in first strand etc. that
inherited the properties of the class Nucleotide.

The project is mostly written in a general form and can be applied
to all kinds of faces (hairpin loops, bulges etc.) with slight software
changes.

In order to to have all the initial information of the interior loop,
a PDB file was read (1AFX.pdb) and the information of the nucleotides
comprising the interior loop was stored.

The bases' atoms, throughout the program, were generated using two
base vectors that span the 2d plane in which the base resides (since
the base is rigid and planar). Since the number of configurations
over which we sum their energy values is huge, Kahan's algorithm was
used to reduce accumulative errors of the floating point numbers \citep{Kahan}.

In order to enable us to change the parameters needed for the runs
without having to compile the software each time, parameter files
were written and read in run time.

\subsubsection{Optimization}

As the computational power needed in the simulation is big, optimization
of the running time was taken seriously.

It was done according to the instructions in \citep{C++} with the
aid of Visual Studio performance wizard, which enabled us to identify
the bottlenecks and the time consuming parts of the program. The code
was revised in many places and the running time was minimized mostly
due to the following steps \label{Optimization}:
\begin{itemize}
\item Avoidance of using library functions for calculations if their full
functionality isn't used (factor of $\sim$5 on the total running
time).
\item Avoidance of memory allocations in places where it's not obligatory
(factor of $\sim$5 on the total running time).
\item Use of inline functions in cases in which they are used frequently.
Inlined functions are copied into place instead of being called (factor
of $\sim$2 on the total running time).
\end{itemize}
We then switched to using Intel compiler for c++ (icpc), which optimized
the compilation for the specific processor we were using (factor of
$\sim$2 on the total running time). Moreover, the capabilities of
the new Intel processors to vectorize operations were used, and functions
that were called frequently were re-written in order to meet the requirements
for vectorization (factor of $\sim$3 on the total running time).

Overall the running time of the software was minimized by an overall
factor of $\sim$250, which enabled us to perform simulations in reasonable
amounts of time on the machine and processor we used.

\subsubsection{Visualization of the simulation}

In order to visualize the simulation, the capability to write the
information of the oligonucleotide instance to a PDB file was added.
Since we also wanted to visualize the development of the simulation
in time, PYMOL which is a software that generates a video from PDB
files, was used. PYMOL commands have been studied in order to enable
us to create movies. Then, the commands were integrated into the software
in order to generate automatically a script file that, when double
clicked, shows a movie of the PDB files.

\subsubsection{Simulation details}

The simulation for each of the four systems consisted of parallel
tempering over five temperature ranges (see Appendix \eqref{sub:Temp-chosen}),
each performed on a single core. For each replica there were 200,000
configuration exchange attempts with replicas at adjacent temperatures.
Since there are $\sim1000$ Monte-Carlo moves between adjacent configurations
exchange attempts, there were $\sim2\cdot10^{8}$ Monte Carlo moves
for each temperature. Overall, for the four systems for all the temperatures
simulated for the two cutoff energies$\sim64\cdot10^{9}$ Monte-Carlo
micro-states were generated, from which a movie of a length of $\sim68$
years can be created (30 FPS).

\newpage{}

\subsection{Results}

\subsubsection{Consistency checks}

The consistency of the results was checked in several ways:
\begin{itemize}
\item In the canonical ensemble, the following relation is satisfied: $\left\langle \left(\triangle E\right)^{2}\right\rangle =-\frac{\partial U}{\partial\beta}$.
Since in the simulation we can calculate the variance of the energy
and also the internal energy as a function of beta, we were able to
plot and compare the two sides of the equation. The results for the
different bases are shown in figure \ref{fig:Variance-of-the-energy-partial-derivative-of-the-internal-energy}:\smallskip{}
\begin{figure}[H]
\caption{\label{fig:Variance-of-the-energy-partial-derivative-of-the-internal-energy}Variance
of the energy and crude partial derivative of the energy with respect
to beta as a function of temperature {[}(Kcal/Mole)\textasciicircum{}2{]}}

\centering{}\includegraphics[scale=0.18]{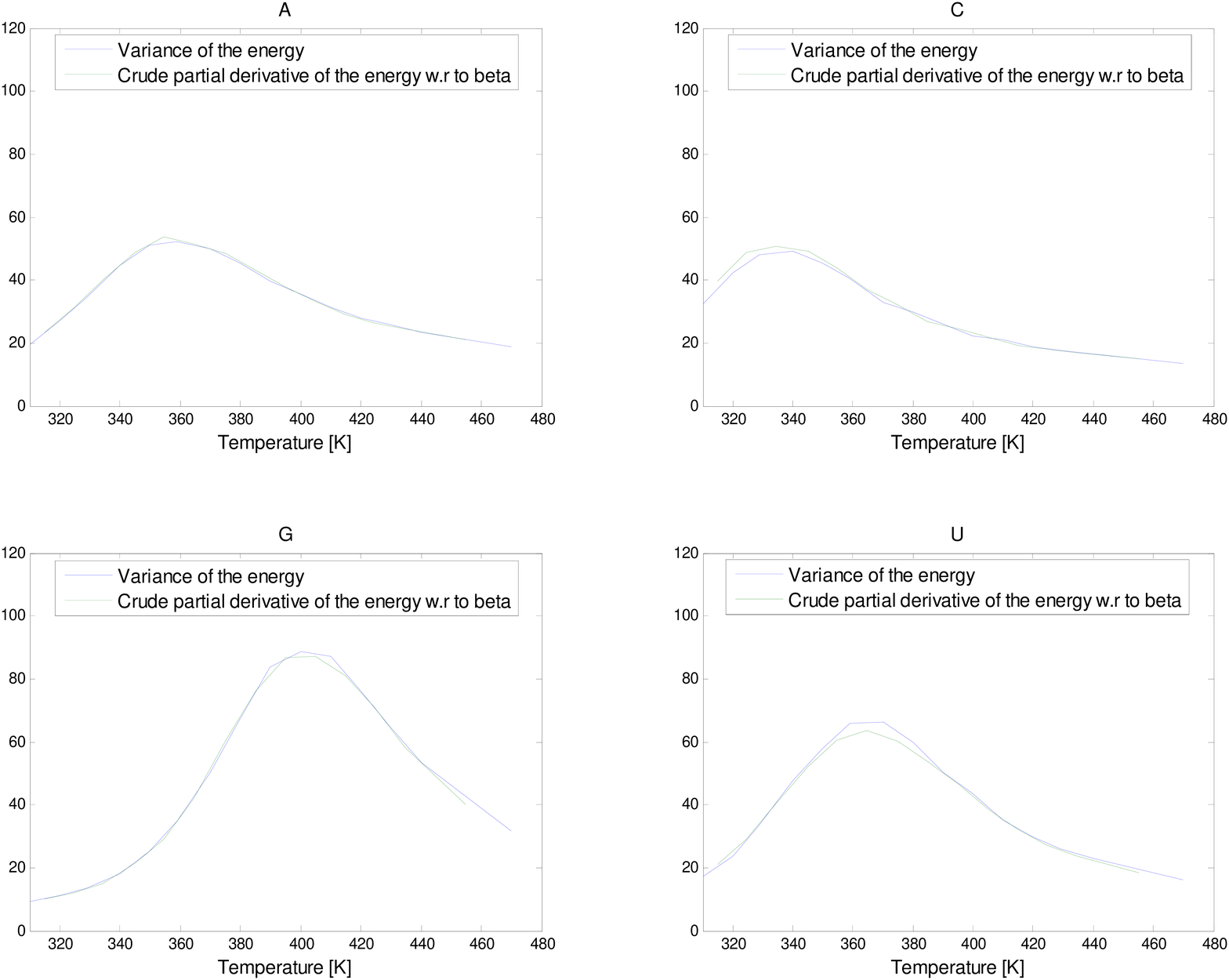}
\end{figure}
It can be seen that the agreement is good, up to deviations that originate
mainly from the relatively large steps in beta and from the higher
standard deviations in the internal energy near the peaks.\newpage{}
\item The energy distribution function at given betas $\beta_{1},\beta_{2}$
are: $P\left(E,\beta_{1}\right)=\frac{g\left(E\right)exp\left(-\beta_{1}E\right)}{Z_{\beta_{1}}}$,
$P\left(E,\beta_{2}\right)=\frac{g\left(E\right)exp\left(-\beta_{2}E\right)}{Z_{\beta_{2}}}$.
Hence, we can estimate $P\left(E,\beta_{2}\right)$ using $P\left(E,\beta_{1}\right)$
as follows: $P\left(E,\beta_{2}\right)_{est}=\frac{P\left(E,\beta_{1}\right)exp\left(-E\left(\beta_{2}-\beta_{1}\right)\right)}{\int P\left(E,\beta_{1}\right)exp\left(-E\left(\beta_{2}-\beta_{1}\right)\right)dE}$.
The results are shown in the following figure:
\begin{figure}[H]
\caption{Energy distribution for T=310K,T=320K, and estimation for T=320K with
U as the second base}

\centering{}\includegraphics[scale=0.16]{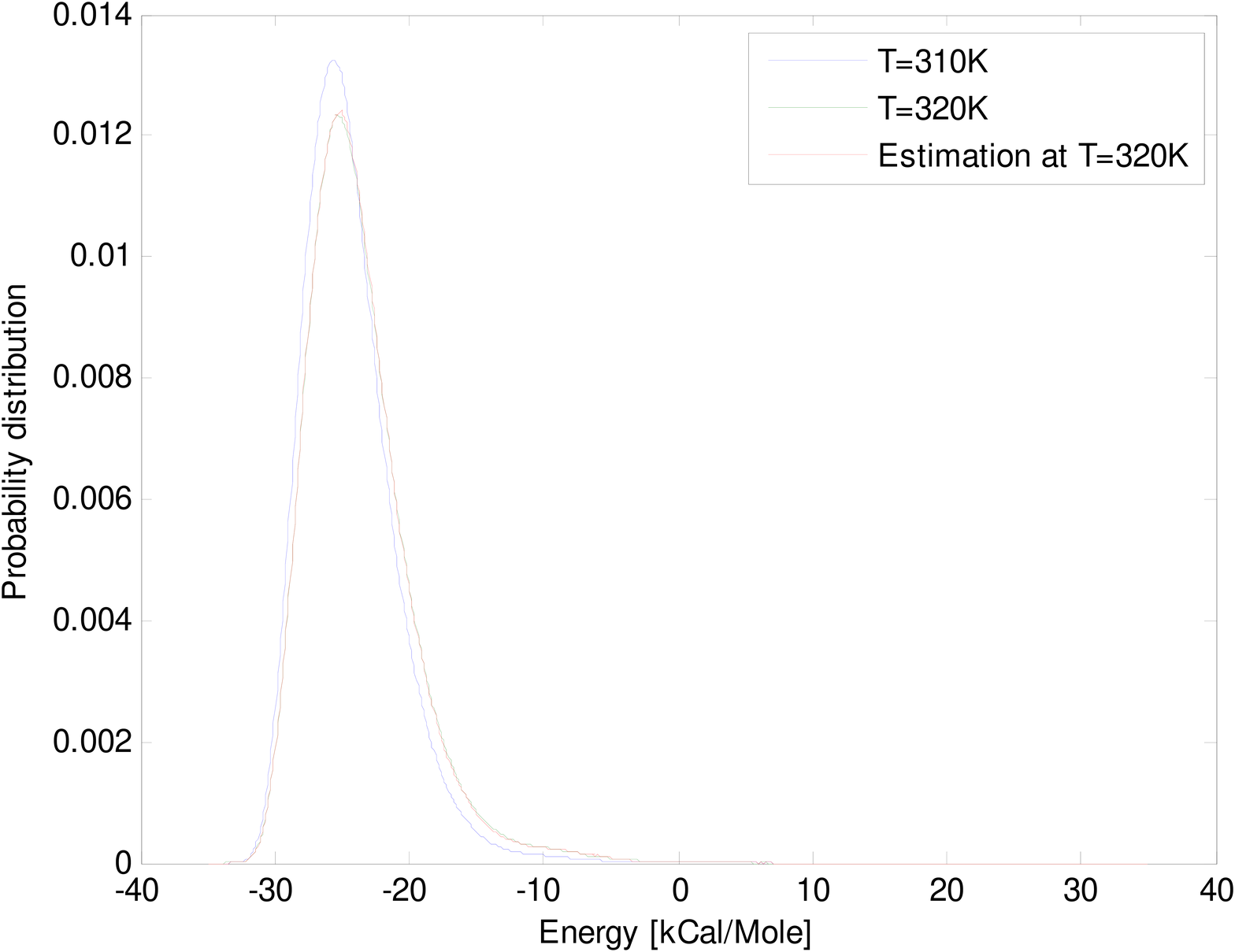}
\end{figure}
 It can be seen that there is good agreement between the energy distribution
at T=320K (green line) and the one estimated at T=320K (red line),
based on the distribution at T=310K (the red and green lines practically
coincide).
\item It had to be verified that the integral $\frac{1}{\beta_{1}}\intop_{0}^{\beta_{1}}\left\langle H_{X}\right\rangle d\tilde{\beta}$
is independent of the cutoff energy as discussed in \ref{sub:cutoff},
and the results are shown later on.
\end{itemize}

\subsubsection{Energetically favorable micro-states}

We ran the simulation and captured the energetically favorable micro-states
in order to asses the energy model. The selected micro-states, and
their energy values for the configuration with different bases are
presented in the following figures: (1-3 first strand, 4-6 second
strand. atoms by their color: white - Hydrogen, blue- Nitrogen, red-
Oxygen, green - Carbon)

\begin{figure}[H]

\caption{Micro-state of the system with base A with low energy. Energy values
(kCal/Mole): Total -23.12, Stacking -20.46, Hydrogen -5.19, Electrical
1.99, Steric 0.54}

\centering{}\includegraphics[scale=1]{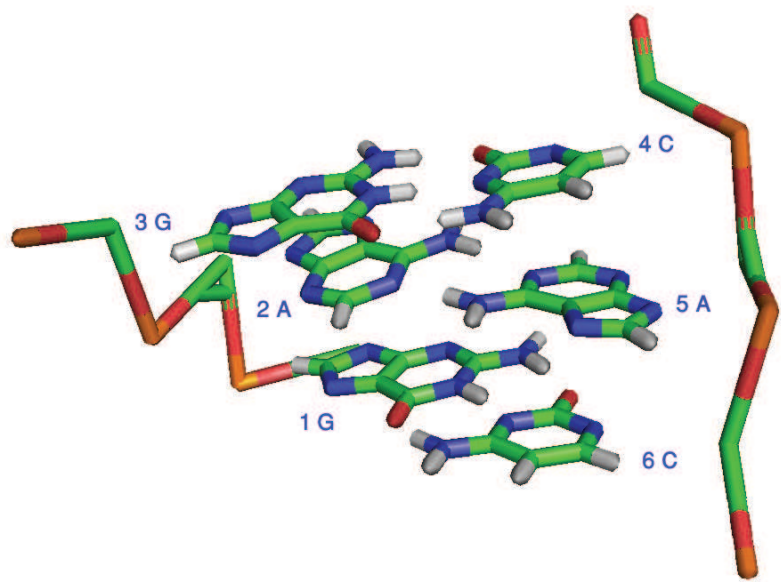}
\end{figure}

\begin{figure}[H]
\caption{Micro-state of the system with base C with low energy. Energy values
(kCal/Mole): Total -23.05 , Stacking -20.38, Hydrogen -5.29 , Electrical
1.9 , Steric 0.72}

\centering{}\includegraphics[scale=1]{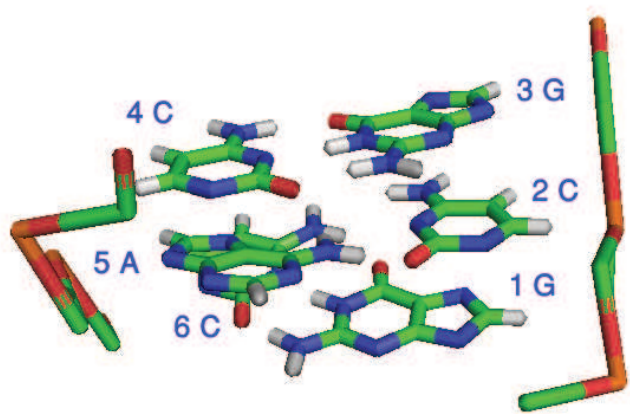}
\end{figure}

\begin{figure}[H]

\caption{Micro-state of the system with base G with low energy. Energy values
(kCal/Mole): Total <23 , Stacking -20.42 , Hydrogen <-8.5, Electrical
2.09, Steric 3.29 }

\centering{}\includegraphics[scale=1]{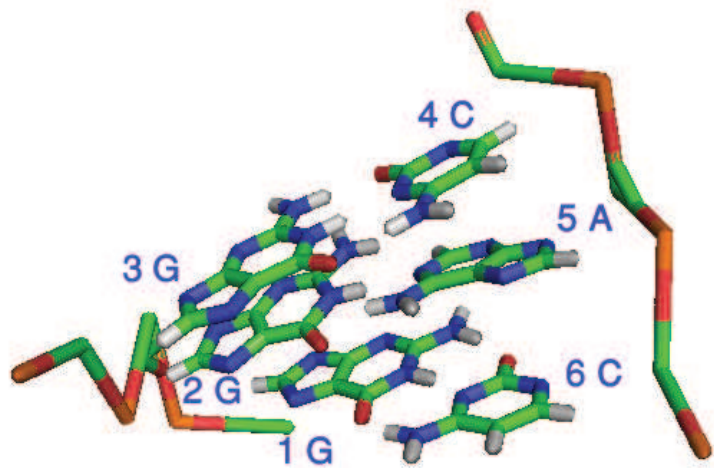}
\end{figure}

\begin{figure}[H]
\caption{Micro-state of the system with base U with low energy. Energy values
(kCal/Mole): Total -23.04, Stacking -16.78, Hydrogen -8.63, Electrical
1.88, Steric 0.5}

\centering{}\includegraphics[scale=1]{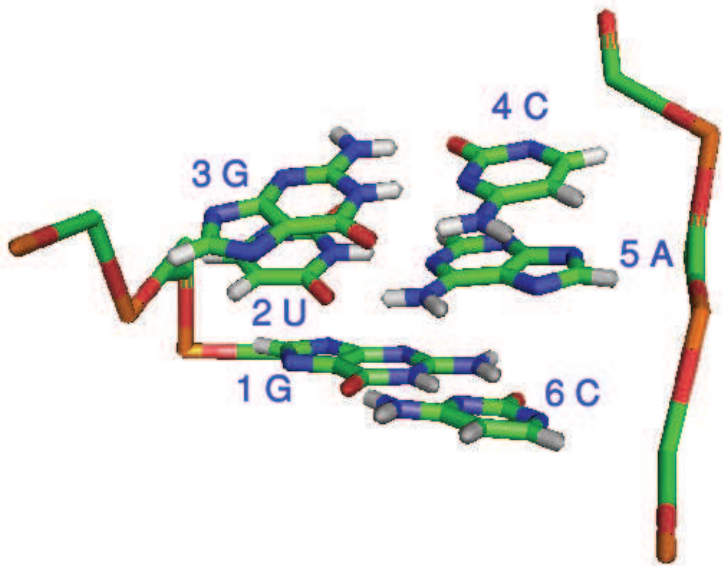}
\end{figure}

It can be seen that the geometries of the 2nd base with respect to
the 5th base in the low energy states are similar to the ones reported
in the literature. The A-G pairing geometry is similar to the one
presented in \citep{leontis2001geometric}. The A-A pairing geometry
is also similar to the one presented in \citep{leontis2001geometric}
and involves Hoogsteen base pairing. The U-A pairing geometry is similar
to the well known Watson-Crick Base pairing.

\subsubsection{Acceptance rates}

The acceptance rates of the internal and external moves as a function
of temperature are shown in the following figure:

\begin{figure}[H]
\caption{Acceptance rate of internal and external moves - system with U as
the second base}

\begin{centering}
\includegraphics[scale=0.15]{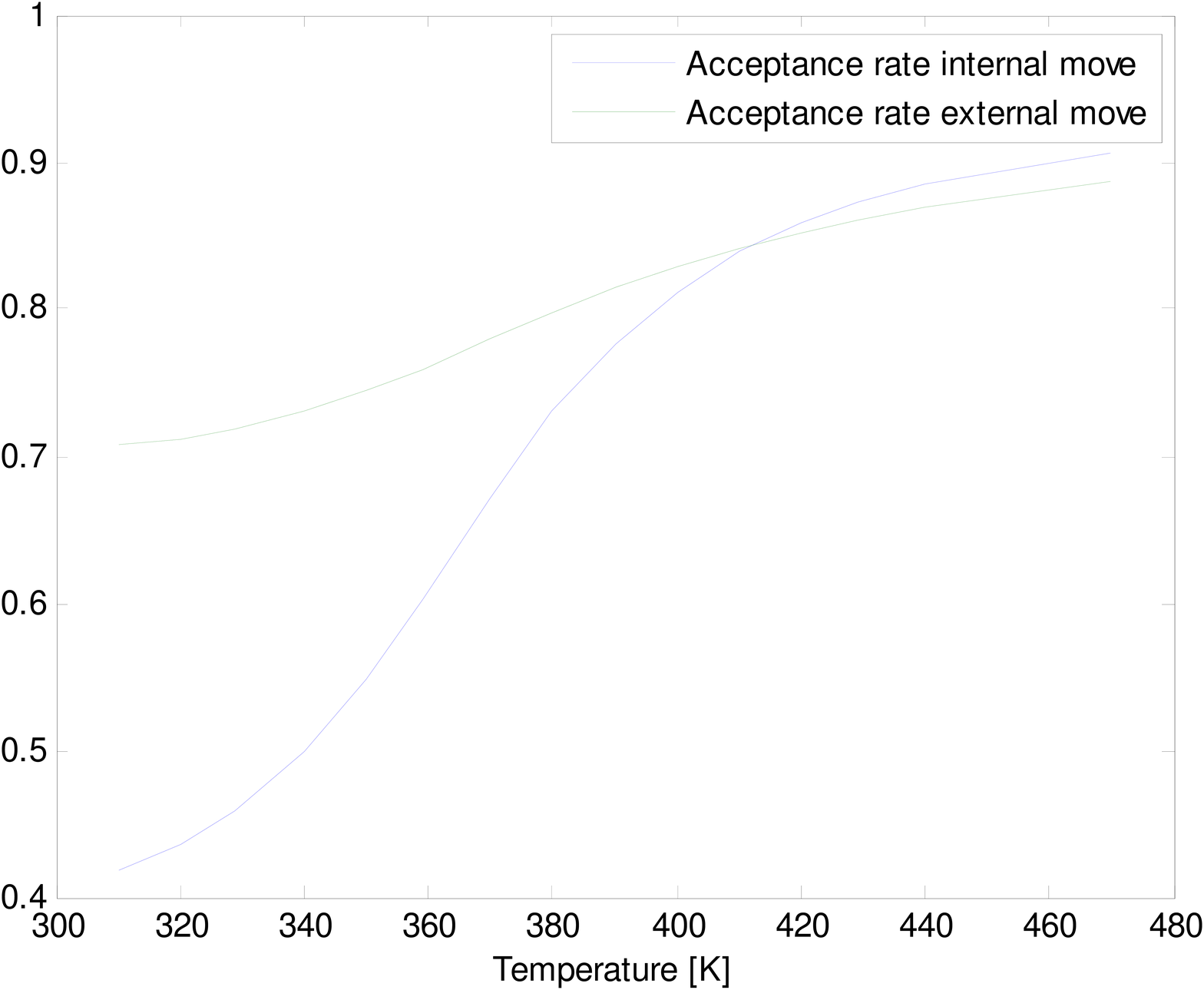}
\par\end{centering}

\end{figure}

As expected, these acceptance rates are larger for higher temperatures.

The configuration exchange acceptance rates were high since we took
small steps in temperature in order to sample well all the temperature
range (values around$\sim0.9$).

\newpage{}

\subsubsection{Results}

Here we present the average of the total, stacking, Hydrogen bond,
electrical, and steric energies for the four systems:

\begin{figure}[H]
\caption{The different average energies as a function of temperature for system
with the second base A,C}

\centering{}\includegraphics[scale=0.201]{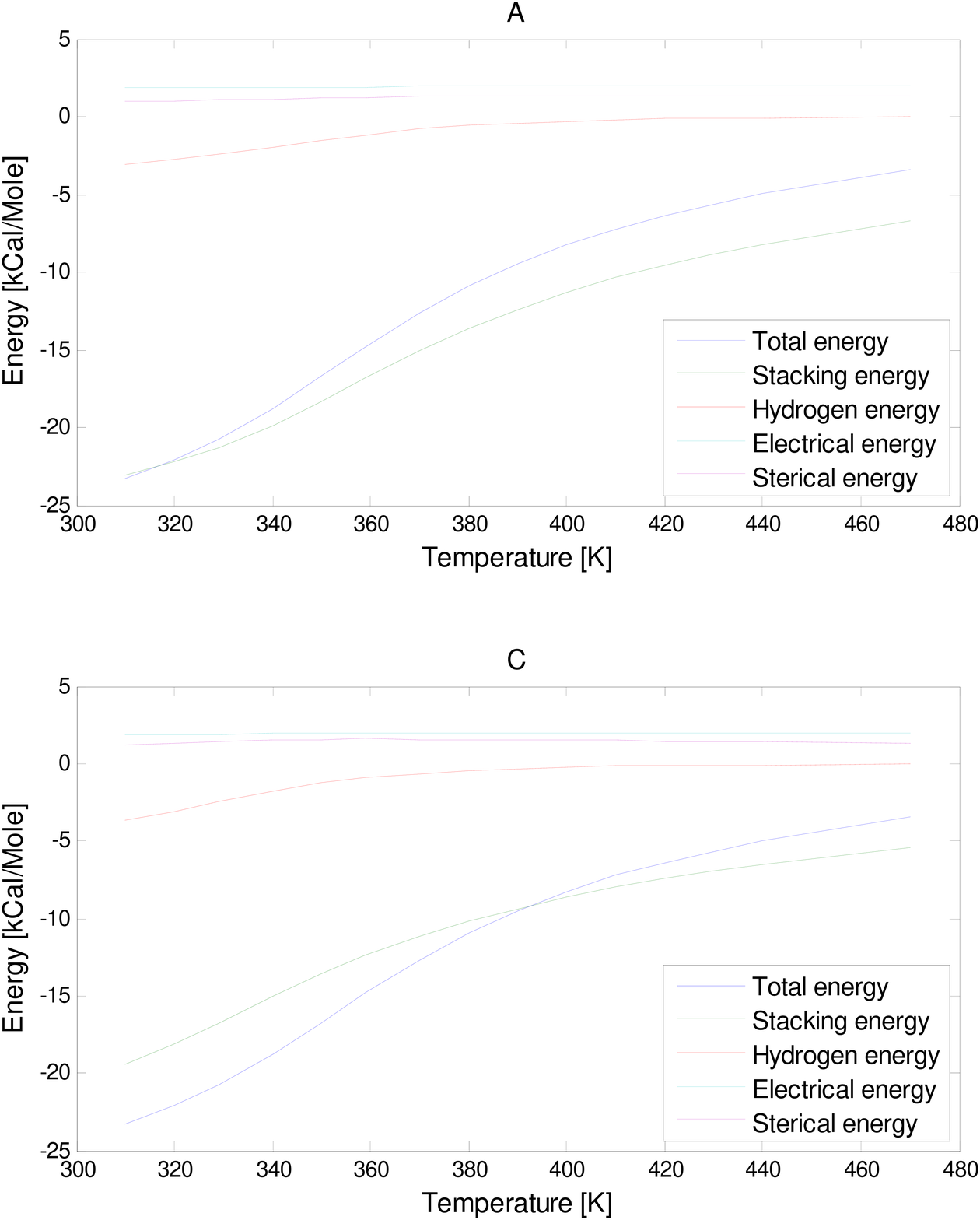}
\end{figure}

\begin{figure}[H]

\caption{The different average energies as a function of temperature for system
with the second base G,U}

\begin{centering}
\includegraphics[scale=0.201]{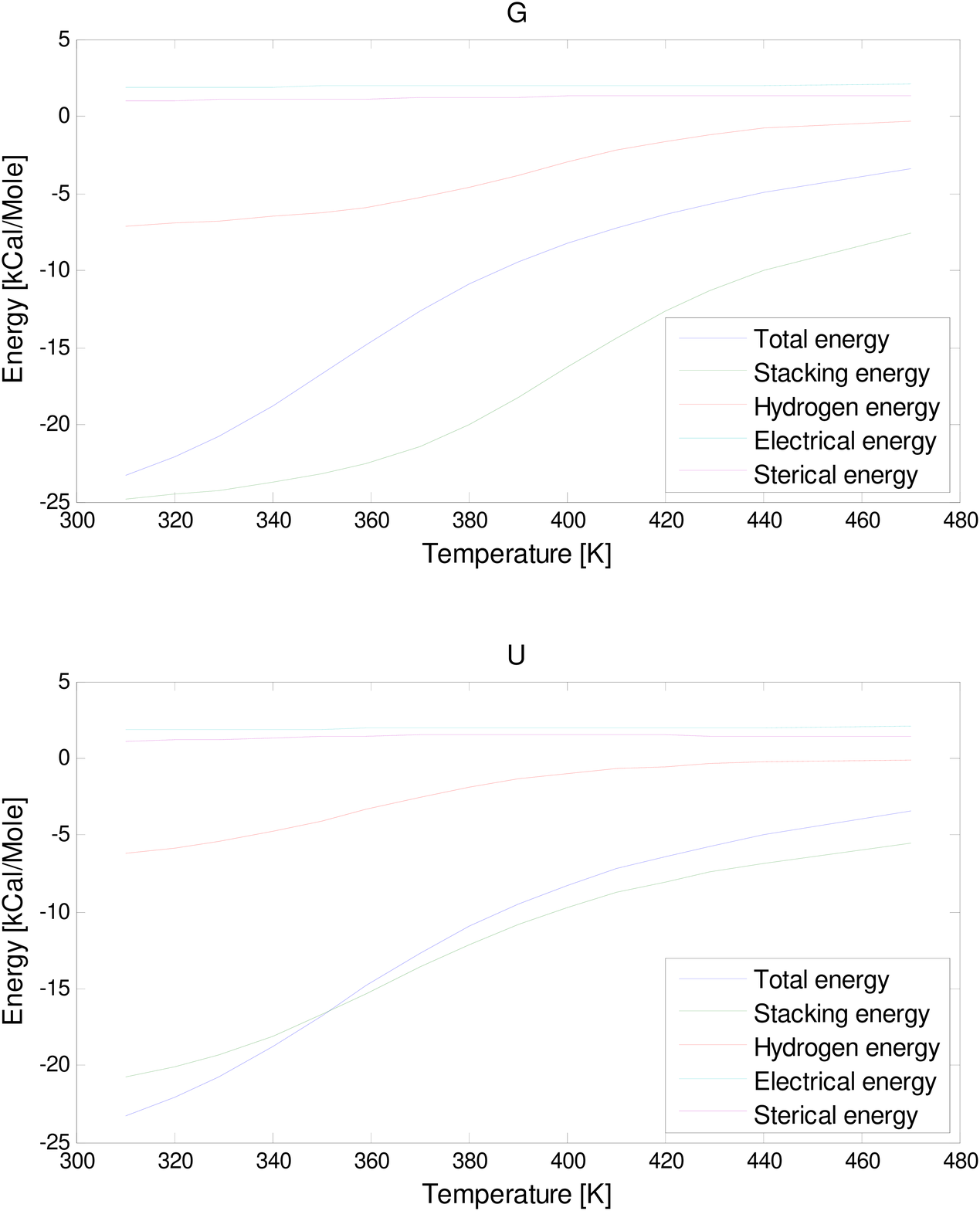}
\par\end{centering}

\end{figure}

It can be seen that the dominating terms are the stacking and Hydrogen
interaction (the Hydrogen interaction energy of the first and last
pair aren't taken into account). As expected all the energies are
monotonically increasing function of the temperature. The most significant
change is in the stacking and Hydrogen energies since there's unwinding
of the pair in the middle. \newpage{}The following is a graph of
the heat capacity at constant pressure $C_{P}=\left(\frac{\partial H}{\partial T}\right)_{N.P}$
for the four systems:

\begin{figure}[H]
\caption{$C_{p}$ as a function of temperature for the different bases}

\begin{centering}
\includegraphics[scale=0.2]{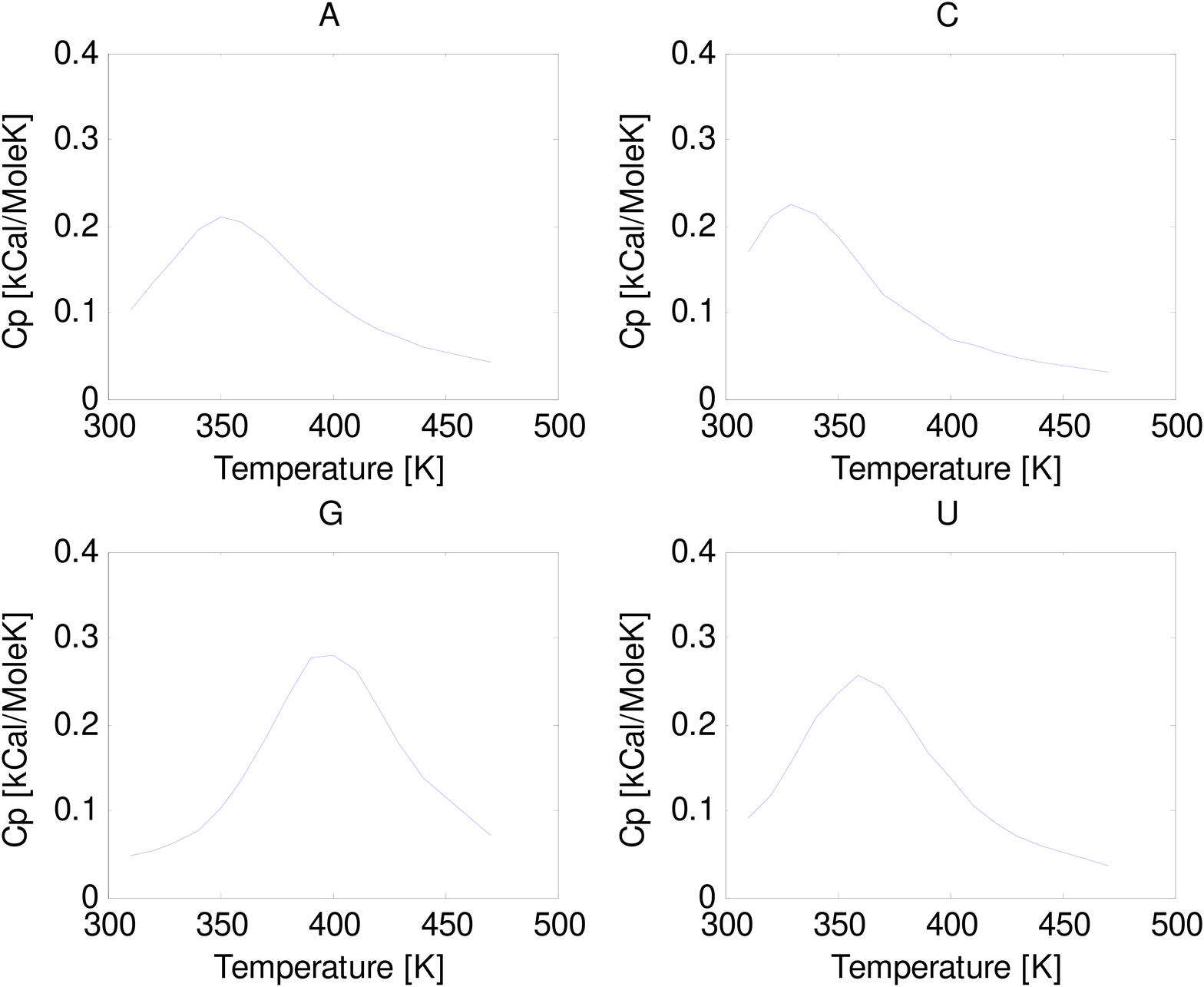}
\par\end{centering}

\end{figure}

In all the configurations there is a maximum in the heat capacity.
This maximum represents the transition from a system in which the
dominant energy terms are the stacking and Hydrogen energies to a
system which is dominated mainly by the steric energy terms.

\newpage{}

Here we present the normalized energy histogram for various temperatures
for a system with the second base U:

\begin{figure}[H]
\caption{Normalized energy histogram for different temperatures for a system
with the second base U}

\centering{}\includegraphics[scale=0.1]{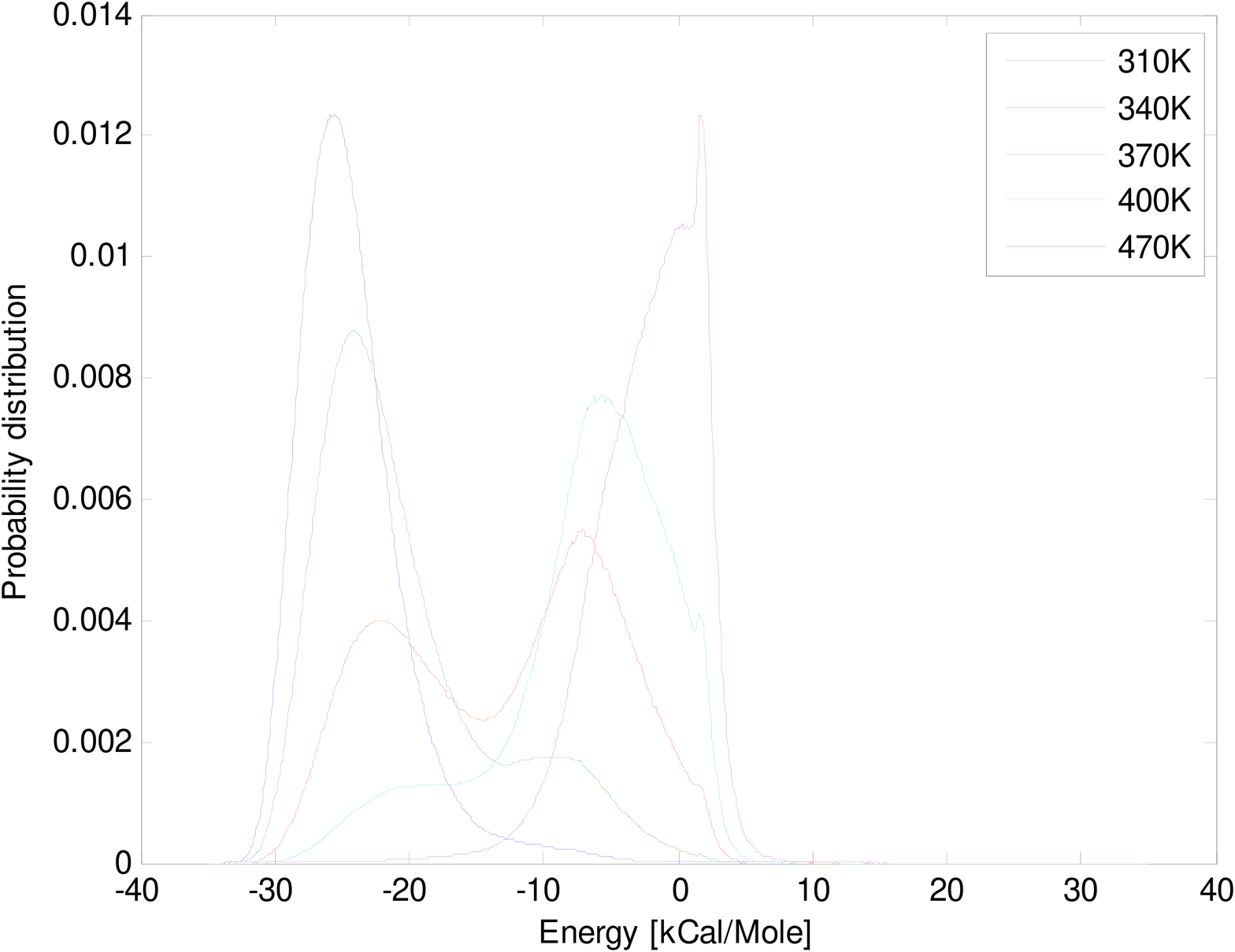}
\end{figure}

The probability for a given energy for the continuous case is:

$P\left(E\right)=g\left(E\right)e^{-E/kT}/Z$

and in the MC simulation it's simply the normalized histogram of the
energies.

It can be concluded from the graph that there are two peaks in the
density of states at these temperatures. This is since there is an
increase in the energy distribution that can't be attributed to $e^{-E/kT}$
. As a result, the standard deviation in the energy at temperatures
around 370K is high. The following graphs are the internal energies
as a function of $\beta$ :

\begin{figure}[H]
\caption{Internal energy as a function of beta for the different bases}

\begin{centering}
\includegraphics[scale=0.1]{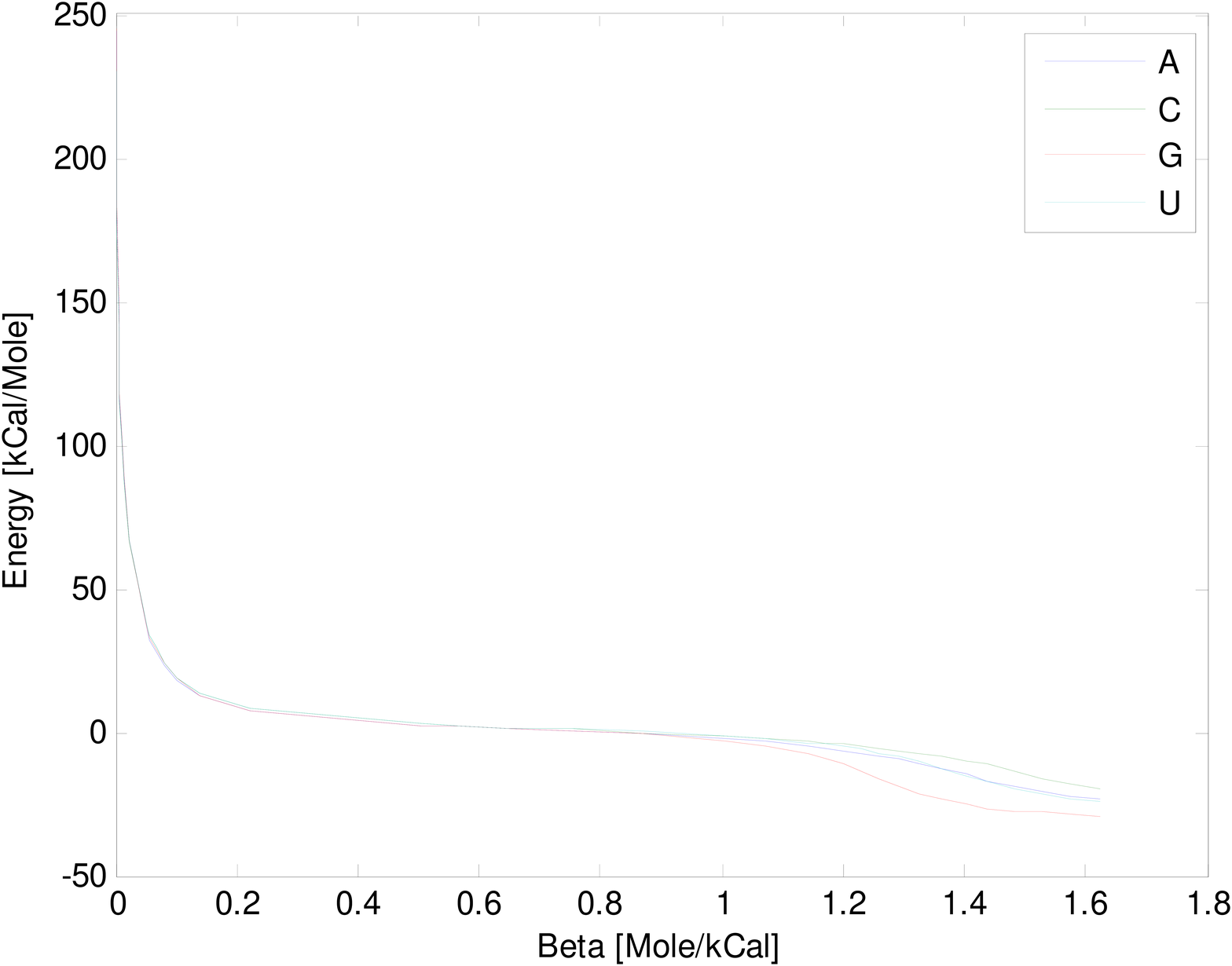}
\par\end{centering}

\end{figure}

This graph is of high importance since it is being integrated over
in the calculation of free energy. The areas differ mainly due to
the behavior in the lower temperature regime. The high energy values
seen at small $\beta s$ appear because the high temperature enables
the molecules to cross one another so the average steric energy is
high. In the next figure we compared the energy as a function of $\beta$
for configurations with bases A and C.

\begin{figure}[H]
\caption{Log of the energy as a function of log $\beta$ - comparison between
interior loops with A and C}

\centering{}\includegraphics[scale=0.1]{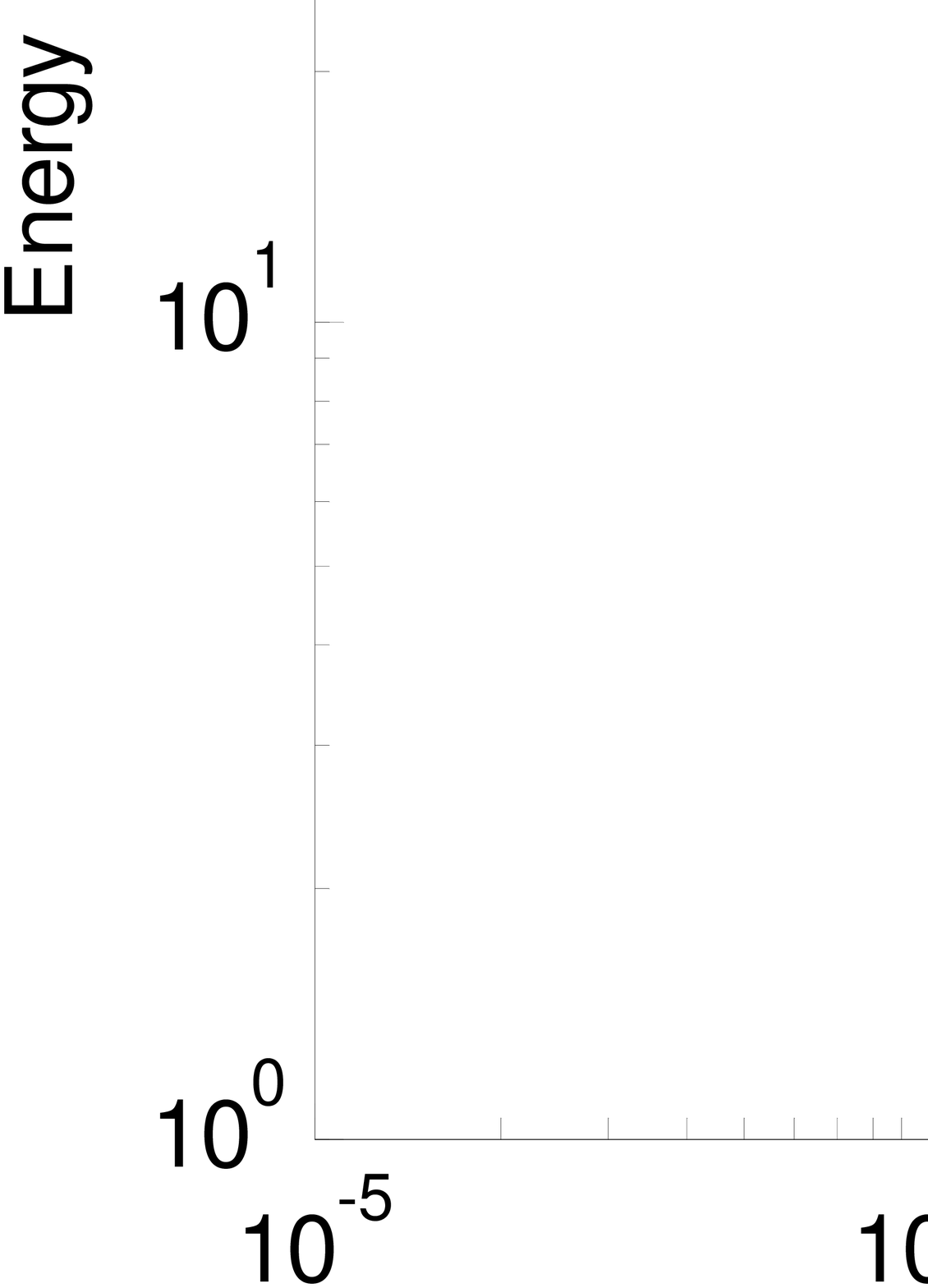}
\end{figure}

As can be seen from the figure, there is saturation in the energy
at small $\beta$s. This is due to the fact that kT at these temperatures
is much higher than the energy terms and there is free movement. The
fact that the configuration with A has higher internal energy at low
$\beta$s can be explained by the higher number of atoms that A contains
compared to C, which results in more atoms on average that have high
energy terms.

As we raise the cutoff energy, the temperature from which the behavior
of the system differs from the original one is higher. Together with
it, the internal energy that we'll have at infinite temperatures will
be higher as the molecules are free to enter the volume of one another
and will experience higher energies on average (since the cutoff energy
is higher). This is in agreement with the assumption that the integral
of the internal energies over beta is independent of the cutoff energies
as long as we satisfy the relevant conditions. In the following graph
we can see that for the higher cutoff we have lower internal energies
at $\beta\sim1\left[Mole/kCal\right]$ and higher internal energies
for $\beta<10^{-2}\left[Mole/kCal\right].$ This behavior is more
discernible as the difference between the cutoff energies becomes
larger.

\begin{figure}[H]
\caption{Log of the energy as a function of log of beta for the two cutoffs
- A}

\centering{}\includegraphics[scale=0.1]{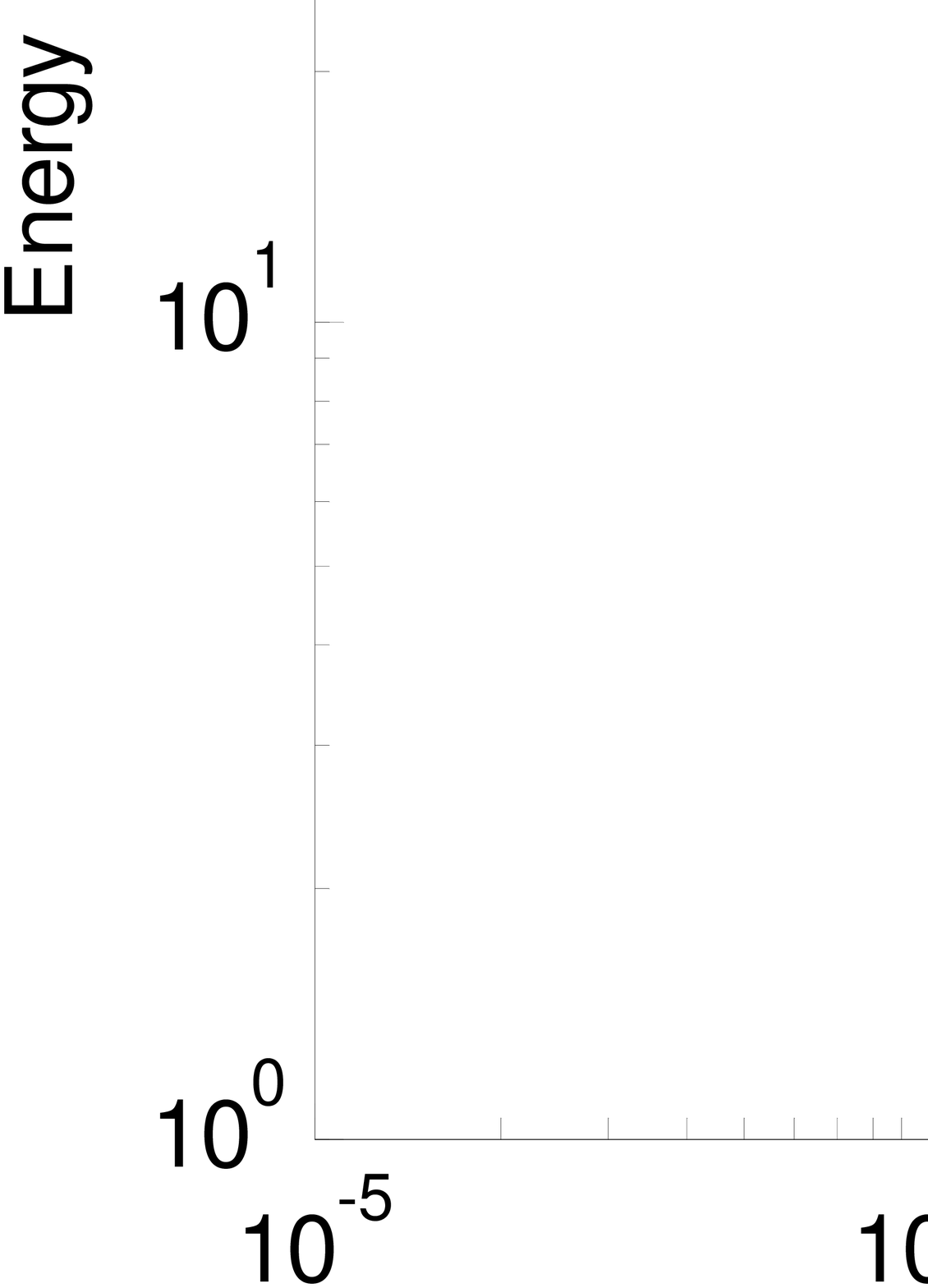}
\end{figure}

We calculated the values of the integral of the internal energy over
$\beta$ for the chosen cutoff energies. The results are shown in
the following figure :
\begin{figure}[H]
\begin{centering}
\caption{Values of the integral for the different cutoff energies (in kCal/Mole)
at body temperature }

\par\end{centering}

\smallskip{}

\centering{}%
\begin{tabular}{|c|c|c|}
\hline
 & $E_{\mathrm{cutoff_{1}}}=5kCal/Mole$ & $E_{\mathrm{cutoff}_{2}}=4.4kCal/Mole$\tabularnewline
\hline
\hline
$\frac{1}{\beta_{1}}\intop_{0}^{\beta_{1}}\left\langle H_{A}\right\rangle d\tilde{\beta}$ & 0.73 & 0.83\tabularnewline
\hline
$\frac{1}{\beta_{1}}\intop_{0}^{\beta_{1}}\left\langle H_{C}\right\rangle d\tilde{\beta}$ & 2.22 & 2.26\tabularnewline
\hline
$\frac{1}{\beta_{1}}\intop_{0}^{\beta_{1}}\left\langle H_{G}\right\rangle d\tilde{\beta}$ & -1.72 & -1.66\tabularnewline
\hline
$\frac{1}{\beta_{1}}\intop_{0}^{\beta_{1}}\left\langle H_{U}\right\rangle d\tilde{\beta}$ & 1.11 & 1.01\tabularnewline
\hline
\end{tabular}
\end{figure}

The integration for the two cutoffs yielded similar results and the
differences aren't far from being in the expected range of error (0.071kCal/Mole,
see appendix \ref{sub:Standard-deviation-in-calculation-of-free-energy}
for details).

As explained in \ref{sub:The-method} the difference between the integrals
are the free energy difference between the configurations, meaning
 that the calculated values are the free energy values of the configurations
up to a constant.

\begin{figure}[H]
\caption{\label{fig:Internal-energies-for-the-different-configuraions}Internal
energies for the different configurations}

\smallskip{}

\centering{}%
\begin{tabular}{|c|c|c|}
\hline
 & $E_{\mathrm{cutoff_{1}}}=5kCal/Mole$ & $E_{\mathrm{cutoff}_{2}}=4.4kCal/Mole$\tabularnewline
\hline
\hline
$U_{A}\left(T=310K\right)$ & -23.29 & -23.23\tabularnewline
\hline
$U_{C}\left(T=310K\right)$ & -20.04 & -19.85\tabularnewline
\hline
$U_{G}\left(T=310K\right)$ & -29.08 & -28.97\tabularnewline
\hline
$U_{U}\left(T=310K\right)$ & -24.01 & -23.98\tabularnewline
\hline
\end{tabular}
\end{figure}

\label{free-energy}Since we have the internal energy of the different
configurations (see Fig \ref{fig:Internal-energies-for-the-different-configuraions}),
we can also calculate the entropy difference.

For example, the entropy difference between $A$ and $G$ is calculated
as follows for $E_{\mathrm{cutoff_{1}}}=5\left[kCal/Mole\right]$:

$\Delta F_{A\rightarrow G}=\Delta H_{A\rightarrow G}-T\Delta S_{A\rightarrow G}=-5.79-310\cdot\Delta S_{A\rightarrow G}=-2.45\left[kCal/Mole\right]\Rightarrow\Delta S_{A\rightarrow G}=-0.0107\left[kCal/KMole\right],-T\Delta S_{A\rightarrow G}=3.34\left[kCal/Mole\right]$

For $E_{\mathrm{cutoff}_{2}}=4.4\left[kCal/Mole\right]$ we have:

$\Delta F_{A\rightarrow G}=\Delta H_{A\rightarrow G}-T\Delta S_{A\rightarrow G}=-5.74-310\cdot\Delta S_{A\rightarrow G}=-2.49\left[kCal/Mole\right]\Rightarrow\Delta S_{A\rightarrow G}=-0.0105\left[kCal/KMole\right],-T\Delta S_{A\rightarrow G}=3.25\left[kCal/Mole\right]$

It can be seen that there is $\sim2\%$ deviation in the entropy value
between the systems with the two cutoffs. This deviation can be minimized
by using higher cutoffs (the cutoffs satisfy $E_{\mathrm{cutoff_{1}}}\approx7kT_{body},E_{\mathrm{cutoff_{2}}}\approx8kT_{body}$)
and larger number of iterations in the MC simulations.

We can also calculate the entropies for higher temperatures in the
same manner. Results are shown in the following figure:

\begin{figure}[H]
\caption{$\Delta S_{A\rightarrow G}$ as a function of temperature}

\begin{centering}
\includegraphics[scale=0.1]{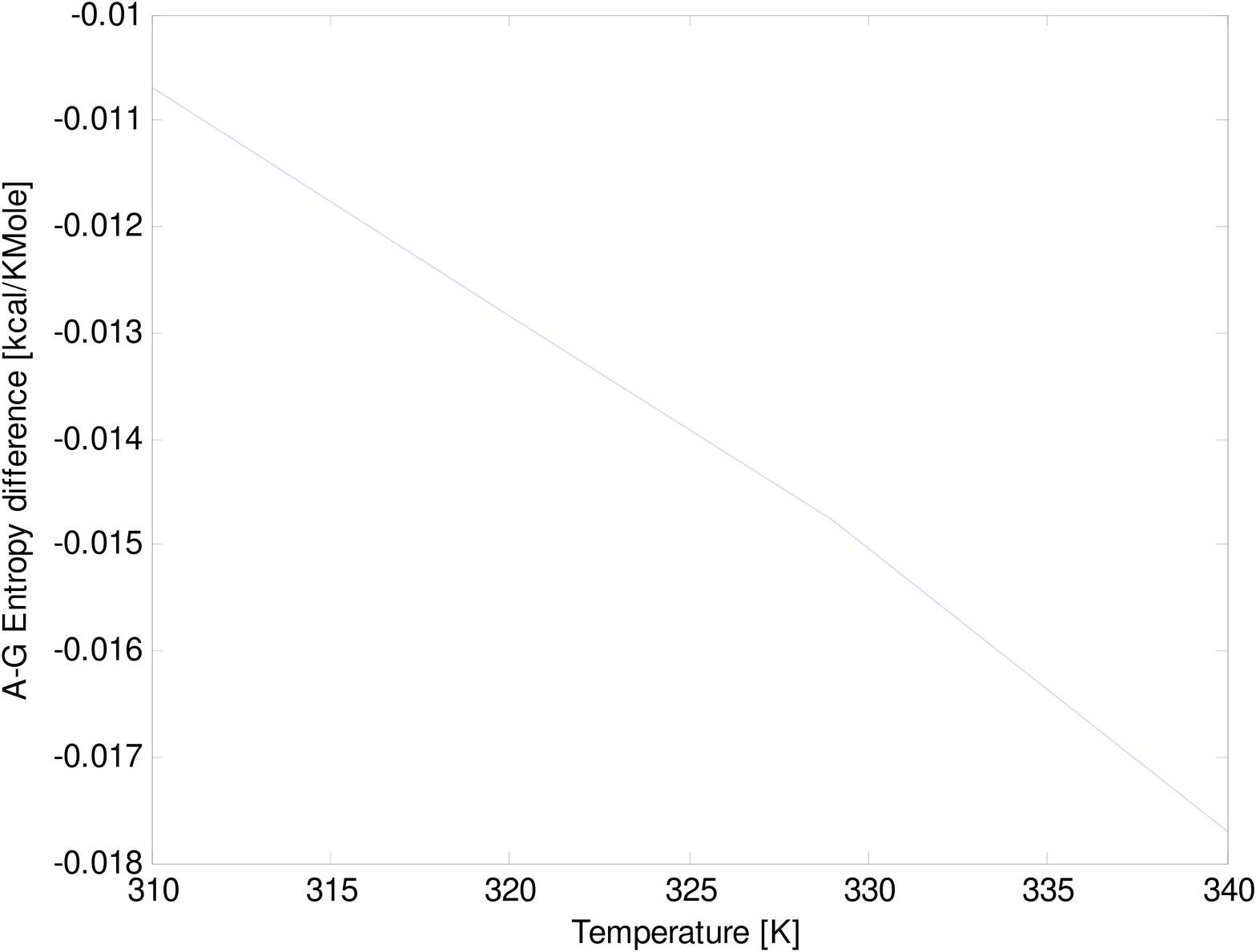}
\par\end{centering}

\end{figure}

The entropy difference in absolute value gets larger as we raise the
temperature. The entropy is a monotonically increasing function of
the internal energy, which is a monotonically increasing function
of the temperature. Hence, as we increase the temperature, the entropies
are expected to get larger and it's likely that the entropy difference
between the systems will also be such.

\subsection{Discussion}

In conclusion, we presented models for faces and the interactions
in RNAs and introduced a procedure to calculate the free energy of
interior loops, and faces in general. This procedure, if verified
or improved may suggest an alternative to the experiments in which
free energy of faces are calculated. Under the assumptions that the
secondary structure is known and that faces are decoupled, simulations
on several faces can also be performed together to generate the whole
tertiary structure, and more global properties can be collected. In
the calculation of free energy we showed a method in which the information
gathered in the course of parallel tempering can be used to calculate
the free energy difference between systems. In this method, if the
energy landscape is rugged, instead of sampling both over the space
of $\lambda$, the parameter related to the TI procedure, and the
temperature, we sample just over the temperatures of the configurations
of interest (up to infinity). This method is much easier to implement
than ThI and is likely to be much faster in rugged energy landscapes.
This is crucial when the cost of the calculation is high, as is often
the case in RNA tertiary structures simulations. In the method we
had to bring the two systems to regimes at which the partition functions
are equal. Although this is easily satisfied for systems that have
the same $\beta\rightarrow0$ limit, we wanted the condition to be
satisfied for all systems that have the same degrees of freedom. We
introduced a cutoff energy in the repulsion term which enabled the
molecules to enter the {}``volume'' of one another and as a result
the condition could be satisfied. This use cutoff energy can also
be applied to other methods such as ThI and enable the calculation
of free energy in the cases where the sampling of the integrand isn't
feasible.

Regarding the optimization of software written in c/c++ languages,
though the range of possible actions is huge, if the bottlenecks are
identified and the relevant actions are performed (as listed in \ref{Optimization}),
together with the use of the proper compiler, and the capabilities
of vectorization of the processor, a tremendous improvement in the
running time can be achieved in a relatively short time.

\subsubsection{Improvement proposals}

As reported by \citep{malathi1981virtual}, the backbone bond angles
defined in our model are in the range of $90^{\circ}-120^{\circ}$.
The P-C-N bond angle also showed small standard deviation by PDB analysis.
In order to take it into account we can add the following energy terms:

$H_{\mathrm{bond\, angle}}=\underset{i}{\sum}\frac{1}{2kT}\left[\left(\frac{\theta_{PCP_{i}}-\theta_{PCP_{avg}}}{\sigma_{\theta_{PCP}}}\right)^{2}+\left(\frac{\theta_{CPC_{i}}-\theta_{CPC_{avg}}}{\sigma_{\theta_{CPC}}}\right)^{2}+\left(\frac{\theta_{PCN_{i}}-\theta_{PCN{}_{avg}}}{\sigma_{\theta_{PCN}}}\right)^{2}\right]$

Alternatively, we can switch to full atomistic modelling of the backbone
and the ribose. As will become clear later, this modelling will include
7 degrees of freedom per nucleotide for the seven torsion angles.
A CONROT (concerted rotation) Monte-Carlo move can be used, in which
7 consecutive torsion angles are changed keeping the bond angles,
distances and the rest of the configuration fixed \citep{dodd1993concerted,ulmschneider2004monte}.
As the backbone torsion angle $\delta$ is directly correlated to
the ribose dihedral angle $\nu_{3}$ which determines the ribose configuration
\citep{saengerp60}, it seems that the ribose doesn't introduce a
degree of freedom \citep{murray2003rna} other than the torsion angle
$\chi$ that defines the orientation of the base with respect to the
ribose. Since the faces are assumed to be decoupled from one another,
the configuration of the ribose and hence $\delta$ and the torsion
angle $\chi$ have to be constant in the nucleotides that are shared
by two faces. Hence, the degrees of freedom that we have consist of
the 7 torsion angles $\alpha,\beta,\gamma,\delta,\varepsilon,\zeta,\chi$
for the nucleotides that are not shared between faces and 5 torsion
angles $\alpha,\beta,\gamma,\varepsilon,\zeta$ for the nucleotides
that are shared between faces. A more natural choice of division of
the atoms that belong to a nucleotide would be C4' to C4' since the
faces are assumed to be decoupled

The ribose also has $\delta$ dependent energy landscape \citep{saengerp60}
which can be taken into account.

Regarding the Monte Carlo moves, we can perform the following moves:
\begin{itemize}
\item CONROT move in the backbone that won't change $\delta$ of the unshared
nucleotides.
\item Rotation of the $\chi$ torsion angle in the unshared nucleotides.
\item A move that will affect the location of the ribose and the base of
the shared nucleotides and that will keep $\delta_{1},\chi_{1},\delta_{2},\chi_{2}$
constant. Such a move can be the following: we can define virtual
bond $C3'_{1}-C4'_{2}$ that connects the two shared nucleotides.
Since the bond and torsion angles between these atoms are assumed
to be constant, the length of this bond can be considered to be fixed.
Since rotation around the $\varepsilon$ or $\gamma$ torsion angle
rotates the virtual bond, the virtual bond angles $O3'_{1}-C3'_{1}-C4'_{2},\, C3'_{1}-C4'_{2}-C5'_{2}$
can be considered fixed. Now, since the requirements for the CONROT
move are satisfied (fixed bond length and bond angle), the $C3'_{1}-C4'_{2}$
can be considered as a bond in a CONROT move that includes two more
atoms and involves the changing the location of the ribose and the
base of the shared nucleotides. This use of virtual bonds in the CONROT
move is rather general and can be used in other places.
\end{itemize}
\cleardoublepage{}

\section{Appendixes}

\subsection{\label{sub:Description-of-the-algorithm}Description of the algorithm
for re-generating structures}

The software reads the string of '(', '.' and ')' characters from
left to right and processes the information iteratively. In each iteration,
the software runs over the '(' and '.' characters until it reaches
a ')' character. In this process it saves the indexes of the '(' characters
in the {}``bases to be paired'' list. Then it runs over the ')'
characters and pairs them with the bases from the list (saves them
in the {}``paired bases'' list) until it reaches a '(' character.
At this point we have a list of {}``paired bases'' and a list of
{}``bases to be paired'' . Here we stop to define a process in which
a son is generated and all the paired bases are moved to it (which
will be called {}``process'' from now on). We also mention here
that each structure has, as one of its fields, the number of unpaired
bases after the process (The initial value for this field is -1).
If, at this point of the iteration, there are more unpaired bases
in the list, a process is done - a son is generated, and the pairs
are stored in it (we'll call this condition 1). If, it continued the
pairing task with the father (if during the iteration, there are more
')' characters after all the bases have been paired, it will continue
the pairing task with the father $\left(*\right)$) , and the father's
substructure has less unpaired bases, than it had before, it means
that it detected a multiloop (called condition 2). In that case, a
son which is a multiloop is created and all the structure's sons are
moved to it. If a son was generated, in the next iteration, if a '('
is detected first, then a new son is generated, and the iteration
is processed with this son $\left(**\right)$.

For example, for the following string:

$\begin{array}{c}
(....(....)....(...(...)....(....)...)....(....)....)\\
A..B...B'...C..D.D'..E..E'..C'..F..F'..A'
\end{array}$

(Where the letters specify locations in the string).

Using the algorithm specified above we'll have the following steps:

\bigskip{}

\begin{tabular}{|>{\centering}b{0.7in}|c|c|c|c|}
\hline
iteration & 1 & 2 & 3 & 4\tabularnewline
\hline
\hline
starting point  & $A$ & $C$ & $E$ & $F$\tabularnewline
\hline
ending point & $C$ & $E$ & $F$ & $A'$\tabularnewline
\hline
conditions fulfilled during the iteration &  &  & $\left(*\right)$,$\left(**\right)$  & $\left(*\right)$,$\left(**\right)$ \tabularnewline
\hline
list of pairs after the iteration & $BB'$ & $DD'$ & $EE'$ & \tabularnewline
\hline
list of bases to be paired after the iteration & $A$ & $A,C$ & $A$ & \tabularnewline
\hline
condition fulfilled after the iteration & 1 & 1 & 2 & 2\tabularnewline
\hline
 &  &  &  & \tabularnewline
 & \includegraphics[scale=0.028]{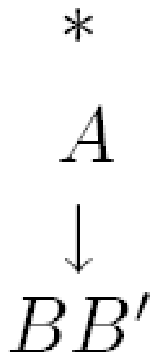} & \includegraphics[scale=0.03]{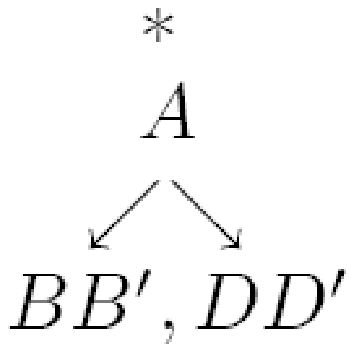} & \includegraphics[scale=0.039]{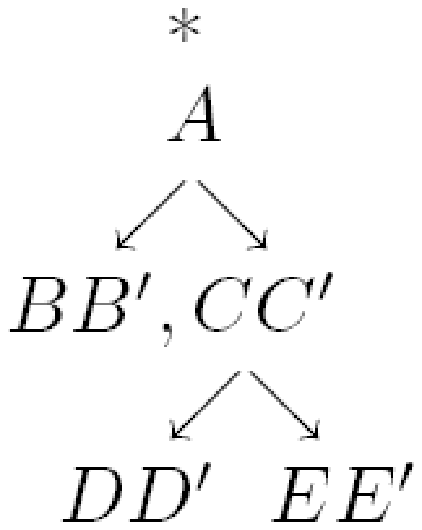} & \multicolumn{1}{c|}{\includegraphics[scale=0.044]{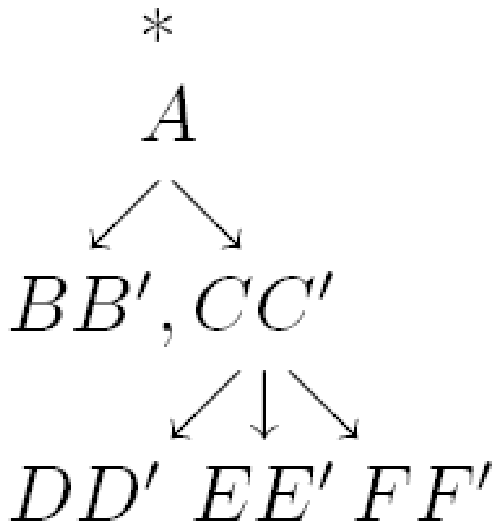}}\tabularnewline
\hline
\end{tabular}

Where {*} is used the specify the structure that the algorithm begins
the iteration with.

The software can give as an output pairs, faces and nucleotide's sequences
for each substructure.

\subsection{\label{sub:Description-of-the-algorithm-comparing}Description of
the algorithm for comparing structures}

For the first structure, it runs over all faces, and for each face,
it looks for identical ones in the second structure. If it finds an
identical face, it updates the faces' field similarity level to identical.
It then removes it from the list of faces of the first and second
structure and adds it to lists of similar faces, in order (identical
face is also similar).

Then it runs over the nonidentical faces in the first structure and
for each face, it looks for similar faces in the second structure.
In case it finds a similar face, it updates the faces' field similarity
level to similar. It then removes them from the list of faces of the
first and second structure and adds it to the new lists. Then it generates
one vector for the first structure and one vector for the second structure
that include the different faces. It then sorts them according to
the face type (hairpin loop, stacking region bulge, interior loop
or multiloop). Thus it enables us to see which faces are preferred
over which faces in the two structures.

\subsection{\label{sub:Experimental-calculation-of-free-energy}Experimental
calculation of free energy}

\subsubsection{\label{sub:General-method}General method}

Say we have two strands of RNA that can bind in the following reaction:

\begin{equation}
\left[A\right]+\left[B\right]\rightarrow\left[AB\right]
\end{equation}

where $\left[A\right],\left[B\right]$ denote the concentrations of
the single strands and $\left[AB\right]$ denotes the concentration
of their bound product.

The free energy difference is:

\begin{equation}
\triangle G=-RTlnK\label{eq:G}
\end{equation}

Where the dissociation constant $K$ is defined as follows:

\begin{equation}
K\triangleq\frac{\left[AB\right]}{\left[A\right]\left[B\right]}
\end{equation}

At $T_{m}$, by definition, the concentration of the reactors and
the products is equal.

Say we take an initial concentration of the reactors of $\left[A\right]=\left[B\right]=\frac{X_{0}}{2}$,
and zero initial concentration of the product $\left[AB\right]=0$.

According to the reaction, it can be seen that at $T_{m}$ (where
$T_{m}$ denotes melting temperature) we have:

$\left[A\right]=\left[B\right]=\left[AB\right]=\frac{X_{0}}{4}$.

Substituting it in \eqref{eq:G}, we get:

\begin{equation}
\triangle G_{T_{m}}=-RTlnK_{T_{m}}=RTln\left(\frac{X_{0}}{4}\right)
\end{equation}

We can use:
\begin{equation}
\triangle G_{T_{m}}=\triangle H\left(T_{m}\right)-T_{m}\triangle S\left(T_{m}\right)
\end{equation}
 and get:

\begin{equation}
\frac{1}{T_{m}}=\frac{R}{H\left(T_{m}\right)}ln\left(\frac{X_{0}}{4}\right)+\frac{\triangle S\left(T_{m}\right)}{\triangle H\left(T_{m}\right)}
\end{equation}

Now each time the initial concentration of the reactors is changed
and $T_{m}$ is measured.

Thus, a graph of $\frac{1}{T_{m}}$ as a function of $ln\left(\frac{X_{0}}{4}\right)$
can be generated.

From this graph $\triangle H\left(T_{m}\right),\triangle S\left(T_{m}\right)$
can be extracted.

It is here to mention that the product consists of many faces and
the values of the enthalpy and entropy are assigned by fitting results
to the faces from many experiments.

\subsubsection{\label{sub:Finding-enthalpy-and-entropy-values-in-body-temperature}Finding
enthalpy and entropy values at body temperature}

Even if we assume that the former experiments are performed in a small
range of temperatures, in which the enthalpy and entropy are constant,
it can't be assumed that these temperatures are close to body temperature.

Measuring the heat capacity as a function of temperature enables the
calculation of the enthalpy and entropy values at body temperature.

\begin{equation}
C_{p}=\left(\frac{dQ}{dT}\right)_{P}
\end{equation}

The enthalpy and entropy can be expressed as functionals dependent
on the function of heat capacity over temperature.

Since the process is quasistatic:

$dQ=TdS$ , and we can write:

\begin{equation}
\left(\frac{dS}{dT}\right)_{P}=\frac{1}{T}C_{P}
\end{equation}

Regarding the enthalpy:

$H\left(S,P,N\right)$ since it's a function of $S,P$ and $N$ and
$P,N$ are held constant we can write the derivative of the enthalpy
with respect to the temperature as follows:

\begin{equation}
\left(\frac{\partial H}{\partial T}\right)_{P,N}=\frac{\partial H}{\partial S}\frac{\partial S}{\partial T}=T\frac{1}{T}C_{P}=C_{p}
\end{equation}

So we can write:

\begin{equation}
\triangle H\left(T_{body}\right)=\triangle H\left(T_{m}\right)+\int_{T_{body}}^{T_{m}}C_{p}dT
\end{equation}

\begin{equation}
\triangle S\left(T_{body}\right)=\triangle S\left(T_{m}\right)+\int_{T_{body}}^{T_{m}}\frac{1}{T}C_{p}dT
\end{equation}

And In conclusion :

$\triangle H\left(T_{m}\right),\triangle S\left(T_{m}\right),C_{p}\left(T\right)\Rightarrow\triangle H\left(T_{body}\right),\triangle S\left(T_{body}\right)$

However, in many calculations, dependency of the heat capacity on
the temperature was neglected, so the experiments are to be redone.

\subsection{Agreement between the Metropolis criteria to detailed balance}

According to the detailed balance criteria, it can be written for
the canonical ensemble:

\begin{equation}
\frac{P\left(i\rightarrow j\right)}{P\left(j\rightarrow i\right)}=\frac{P\left(j\right)}{P\left(i\right)}=\frac{e^{-\beta H_{j}}}{e^{-\beta H_{i}}}=e^{-\beta\left(H_{j}-H_{i}\right)}=
\end{equation}

\begin{equation}
\frac{min\left\{ 1,e^{-\beta\left(H_{j}-H_{i}\right)}\right\} }{min\left\{ 1,e^{+\beta\left(H_{j}-H_{i}\right)}\right\} }
\end{equation}

So for the choice of
\begin{equation}
P\left(i\rightarrow j\right)=min\left\{ 1,e^{-\beta\left(H_{j}-H_{i}\right)}\right\} ,P\left(j\rightarrow i\right)=min\left\{ 1,e^{+\beta\left(H_{j}-H_{i}\right)}\right\}
\end{equation}
 detailed balance is fulfilled.

\subsection{Thermodynamic integration}

Thermodynamic integration is a method to calculate free energy differences
between two systems (in our case they can differ by of a base).

In the method a parameter that transforms the first system to the
second one is introduced.

We can define $A$ as the first system, and $B$ as the second system.

We can write the free energy difference as follows:

\begin{equation}
\triangle F_{A\rightarrow B}\left(\beta_{1}\right)=\intop_{0}^{1}\frac{dF_{\lambda}}{d\lambda}d\lambda=
\end{equation}

Where $\lambda$ is a general parameter. We can substitute $F=-kTlnZ$
and get:

\[
-kT\intop_{0}^{1}\frac{dlnZ}{d\lambda}d\lambda=
\]

\[
-kT\intop_{0}^{1}\frac{1}{Z}\frac{dZ}{d\lambda}d\lambda=
\]

Substituting $Z=\int e^{-\beta H}d\Omega$ we get:

\[
-kT\intop_{0}^{1}\frac{1}{Z}\frac{\int e^{-\beta_{1}H}d\Omega}{d\lambda}d\lambda=
\]

\[
kT\intop_{0}^{1}\frac{1}{Z}\int e^{-\beta_{1}H}\frac{dH}{d\lambda}d\Omega\beta_{1}d\lambda=
\]

\[
\intop_{0}^{1}\frac{\int e^{-\beta_{1}H}\frac{dH}{d\lambda}d\Omega}{Z}d\lambda=
\]

\[
\intop_{0}^{1}\left\langle \frac{dH}{d\lambda}\right\rangle d\lambda=
\]

For a choice of:

\begin{equation}
H=\lambda H_{B}+\left(1-\lambda\right)H_{A}
\end{equation}

We have for

\[
\frac{dH}{d\lambda}=H_{B}-H_{A}
\]

And it can be written:

\begin{equation}
\triangle F_{A\rightarrow B}\left(\beta_{1}\right)=\intop_{0}^{1}\left\langle H_{B}-H_{A}\right\rangle d\lambda
\end{equation}

Or in a more explicit form:

\[
\triangle F_{A\rightarrow B}\left(\beta_{1}\right)=\intop_{0}^{1}\frac{\left\{ \int\left(H_{B}-H_{A}\right)e^{-\beta_{1}\left[\lambda H_{B}+\left(1-\lambda\right)H_{A}\right]}d\Omega\right\} }{\int e^{-\beta_{1}\left[\lambda H_{A}+\left(1-\lambda\right)H_{B}\right]}d\Omega}d\lambda
\]

\subsection{Parameters chosen for the simulation}

\subsubsection{Temperatures chosen for the free energy calculation}

\label{sub:Temp-chosen}In order to perform the integration, the following
temperatures were chosen:
\begin{itemize}
\item range 1: 310 320 329 340 350 359 370 380 390 400 410 420 429 440 470
\item range 2: 500 580 670 780 890 1000
\item range 3: 1000 2285 3571 4857 6142 7428 8714
\item range 4: 10000 22857 35714 48571 61428 74285 87142
\item range 5: 100000 200000 400000 1000000 10000000
\end{itemize}

\subsection{Standard deviations in the measurements}

\subsubsection{Standard deviation in the calculation of the internal energy}

We can estimate the number of steps it takes for two configurations
to be independent, by the average number of steps it takes for a replica
of the system in a given temperature to exchange configuration with
a replica of the system in an adjacent temperature, multiplied by
the square of the number of temperatures. Since the average number
of steps for a configuration exchange is the number of steps after
which we attempt to exchange, divided the acceptance rate, we can
estimate the number of independent measurements in the following way:
$n=\frac{\mathrm{MC\, moves}}{\mathrm{\left(steps\mathrm{\, for\, a\, configuration\, exchange\, attempt/acceptance\, rate}\right)\cdot N^{2}}}$

For a temperature in the first temperature range we can write:

$n=\frac{2\cdot10^{8}\cdot0.9}{910\cdot225}=879$ (where $n$ is the
number of independent measurements).

Since the standard deviation of the average of n independent variables
with a standard deviation $\sigma_{i}$ is:

$\sigma_{A}=\frac{\sigma_{i}}{\sqrt{n}}$, we can use it to estimate
the standard deviation of the internal energy.

For example for the first temperature, we can substitute the standard
deviation of the energy from the simulation and get:

$\sigma_{T}=\frac{\sigma_{L}}{\sqrt{n}}=\frac{4.148}{29.64}=0.14\left[kCal/Mole\right]$

\subsubsection{\label{sub:Standard-deviation-in-calculation-of-free-energy}Standard
deviation in the calculation of the free energy}

In the free energy calculation, we integrate numerically the average
energy over function of beta:

$\frac{1}{\beta_{1}}\left[\intop_{0}^{\beta_{1}}\left\langle H_{X}\right\rangle d\tilde{\beta}\right]$

It is done by estimating the area between two points as the average
energy multiplied by the difference in $\beta$.

$F_{X}=\frac{1}{\beta_{1}}\left[\intop_{0}^{\beta_{1}}\left\langle H_{X}\right\rangle d\tilde{\beta}\right]=\frac{1}{\beta_{1}}\underset{i}{\sum}\left(\left\langle H_{X}\left(\beta_{i}\right)\right\rangle +\left\langle H_{X}\left(\beta_{i+1}\right)\right\rangle \right)/2\cdot\Delta\beta_{i}$

The standard deviation and the error range can be calculated as follows:

$\sigma_{F_{x}}=\frac{1}{2\beta_{1}}\sqrt{\left(\underset{i}{\sum}\sigma_{\left\langle H_{X}\left(\beta_{i}\right)\right\rangle }^{2}+\sigma_{\left\langle H_{X}\left(\beta_{i+1}\right)\right\rangle }^{2}\right)\cdot\Delta\beta_{i}^{2}}=0.0355kCal/Mole$

$\Rightarrow\mathrm{error}\mathrm{\, range\sim2}\sigma_{F_{x}}=0.071kCal/Mole$

\cleardoublepage{}

\bibliographystyle{unsrt}
\bibliography{bib_c}

\end{document}